\documentclass[12pt]{article}

\usepackage{type1cm,amsmath,amssymb,color,graphics,amscd,amsfonts,mathrsfs,epsf,indentfirst}
\usepackage{epsfig}
\usepackage{bm}
\usepackage{subfigure}
\usepackage{multirow,graphicx}
\usepackage{comment}
\usepackage{epstopdf}

\newcommand{\be}{\begin{equation}}
\newcommand{\ee}{\end{equation}}
\newcommand{\bea}{\begin{eqnarray}}
\newcommand{\eea}{\end{eqnarray}}

\title{{\bf On the horizon instability of an extreme Reissner-Nordstr\"om black hole}}

\author{James Lucietti$^a$\footnote{j.lucietti@ed.ac.uk }\ , Keiju Murata$^{b,c}$\footnote{K.Murata@damtp.cam.ac.uk}, Harvey S. Reall$^b$\footnote{hsr1000@cam.ac.uk} \\ and Norihiro Tanahashi$^d$\footnote{tanahashi@ms.physics.ucdavis.edu} \\ \\
\small \sl $^a$  School of Mathematics and Maxwell Institute of Mathematical Sciences, \\ \small \sl University of Edinburgh,  Edinburgh, EH9 3JZ, UK \\
\small \sl $^b$ Department of Applied Mathematics and Theoretical Physics, University of Cambridge,\\ \small \sl Wilberforce Road, Cambridge CB3 0WA, UK \\
\small \sl $^c$ Yukawa Institute for Theoretical Physics, Kyoto University, Kyoto, 608-8502, Japan\\
\small \sl $^d$  Department of Physics, University of California, Davis, CA 95616, USA}

\begin{document}

\maketitle

\begin{abstract}
Aretakis has proved that a massless scalar field has an instability at the horizon of an extreme Reissner-Nordstr\"om black hole. We show that a similar instability occurs also for a massive scalar field and for coupled linearized gravitational and electromagnetic perturbations. We present numerical results for the late time behaviour of massless and massive scalar fields in the extreme RN background and show that instabilities are present for initial perturbations supported outside the horizon, e.g.\ an ingoing wavepacket. For a massless scalar we show that the numerical results for the late time behaviour are reproduced by an analytic calculation in the near-horizon geometry. We relate Aretakis' conserved quantities at the future horizon to the Newman-Penrose conserved quantities at future null infinity.
\end{abstract}

\newpage

\tableofcontents

\section{Introduction}

Supersymmetric black holes play an important role in string theory. Given their importance, it is natural to ask whether or not they are classically stable: does a small initial perturbation remain small under time evolution? 

In a supergravity theory, the fact that a supersymmetric solution saturates a BPS bound, and therefore minimises the energy (at fixed charge), does {\it not} imply classical stability. For example, the classical stability of Minkowski spacetime in vacuum GR does not follow from the positive energy theorem. Instead, the proof involves a lengthy analysis of the Einstein equation \cite{Christodoulou:1993uv}. Furthermore, Anti-de Sitter spacetime is a supersymmetric solution of various supergravity theories but nevertheless it is classically unstable against the formation of small black holes \cite{Bizon:2011gg}. 

In these examples we are referring to classical stability under time evolution determined by the {\it nonlinear} Einstein equation. Recently, Aretakis has proved that even linear perturbations of a supersymmetric black hole can exhibit an instability \cite{Aretakis:2011ha,Aretakis:2011hc}.\footnote{
Ref. \cite{Marolf:2010nd} conjectured the existence of an instability of extreme black holes.} He considered an extreme Reissner-Nordstr\"om black hole. In ingoing Eddington Finkelstein (EF) coordinates, the metric is
\be
\label{extremeRN}
 ds^2 = -\left( 1- \frac{M}{r} \right)^2 dv^2 + 2 dv dr + r^2 d\Omega^2 \; .
\ee
The future event horizon ${\cal H}^+$ is at $r=M$. The generator of time translations is $\partial/\partial v$. This solution preserves half of the supersymmetry in minimal $N=2$ supergravity \cite{Gibbons:1982fy}.

\begin{figure}
\begin{center}
\includegraphics[scale=0.3]{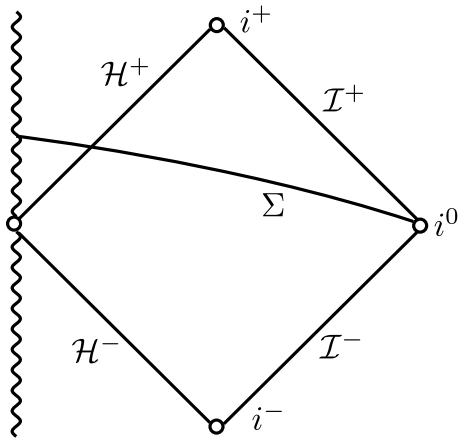}
\end{center}
\caption{
Penrose diagram for extreme RN black holes.
Aretakis took initial data specified on a spacelike surface $\Sigma$
 intersecting ${\cal H}^+$ and extending to infinity.
}
 \label{penrose}
\end{figure}

Aretakis studied a massless scalar field $\psi$ in this spacetime. Consider initial data for $\psi$ specified on a spacelike surface $\Sigma$ intersecting ${\cal H}^+$ and extending to infinity  (Fig.~\ref{penrose}), with $\psi$ decaying at infinity. This uniquely determines $\psi$ in the future domain of dependence of $\Sigma$, which contains 
the part of the black hole exterior that lies to the future of $\Sigma$, and a neighbourhood of the part of ${\cal H}^+$ that lies to the future of $\Sigma$. Aretakis proved that the solution exhibits both stable and unstable features. Stable features are that $\psi$ decays at late time (large $v$) on, and outside, ${\cal H}^+$ and all derivatives of $\psi$ decay outside ${\cal H}^+$. However, by constructing certain conserved quantities on ${\cal H}^+$ (reviewed below), Aretakis proved that $\partial_r \psi$ generically does not decay at late time {\it on} ${\cal H}^+$. This implies the existence of an instability: since $\partial_r \psi$ decays for $r>M$ but not for $r=M$, it follows that $\partial_r^2 \psi$ must blow up at late time on ${\cal H}^+$. Aretakis proved that $\partial_r^k \psi$ generically blows up at least as fast as $v^{k-1}$ at late time on ${\cal H}^+$. 

This instability is not a coordinate effect because $-\partial/\partial r$ can be invariantly defined as the tangent to ingoing radial null geodesics of unit energy. The instability involves polynomial growth in time, which would make it hard to discover using a mode analysis. The above results imply that the component $T_{rr}$ of the energy momentum tensor of $\psi$ decays at late time {\it outside} ${\cal H}^+$ but not {\it on} ${\cal H}^+$. One might regard this as ``hair" on the horizon of the black hole. Note that $T_{rr}$ is closely related to the energy density measured by an infalling observer. 

We emphasise that non-extreme black holes such as Schwarzschild or non-extreme Kerr have been proved to be  stable against massless scalar field perturbations: $\psi$ and all its derivatives decay on, and outside, ${\cal H}^+$ \cite{Dafermos:2008en,Dafermos:2010hd}. One might think that extreme black holes also should be stable because the gravitational backreaction of a perturbation would make the black hole non-extreme. However, backreaction on the metric is a nonlinear effect. It seems unlikely that nonlinear effects would eliminate a linear instability. What {\it might} happen is that the endpoint of the instability is generically a non-extreme black hole, i.e., derivatives of $\psi$ become large along ${\cal H}^+$ (so there is still an instability) but eventually are damped by nonlinear effects. Second, even though generic initial perturbations will make the final black hole non-extreme, there are probably non-generic initial perturbations for which the horizon is extreme at late time.

Minimal $N=2$ supergravity does not contain a scalar field but the Aretakis instability can be embedded in a supersymmetric theory as follows. Consider a supersymmetric static 4-charge black hole solution of type II supergravity compactified on $T^6$. In general, this solution has non-trivial moduli fields but if the 4 charges are set equal then the moduli are all constant and the geometry is that of extreme RN. Linearized fluctuations in the moduli are massless scalars, hence Aretakis' result implies the existence of a linearized instability at the horizon of an extreme RN black hole in this theory.

Aretakis has considered also the case of a massless scalar field $\psi$ in the extreme Kerr geometry. He has proved decay of an axisymmetric field $\psi$ \cite{Aretakis:2011gz} and that derivatives of $\psi$ exhibit an instability at ${\cal H}^+$ analogous to the extreme RN instability \cite{Aretakis:2012ei}. Ref. \cite{Lucietti:2012sf} generalised these results to show that a massless scalar field instability occurs at ${\cal H}^+$ for {\it any} extreme black hole and that a similar instability occurs for {\it linearized gravitational} perturbations of the extreme Kerr solution. Ref. \cite{Murata:2012ct} has extended the latter result to linearized gravitational perturbations of certain higher-dimensional extreme vacuum black holes.

Aretakis' proof of instability (reviewed in section \ref{Analytic}) involves an infinite set of conserved quantities at ${\cal H}^+$, linear in $\psi$. We will call these the {\it Aretakis constants}. We will show in section \ref{CI} that these constants are closely related to a set of conserved quantities at ${\cal I}^+$ (future null infinity): the {\it Newman-Penrose constants} \cite{Newman:1968uj}. Indeed, there is a conformal isometry of the extreme RN geometry which interchanges ${\cal H}^+$ with ${\cal I}^+$ \cite{couch}. This map exchanges the Aretakis constants with the NP constants.

The Aretakis instability is associated to {\it outgoing} radiation at ${\cal H}^+$: the proof of instability requires that the initial data be non-vanishing at ${\cal H}^+$ and, as we will explain below (section \ref{redshift}), the instability is closely related to the absence of a redshift for outgoing photons at ${\cal H}^+$. For string theory applications, it seems more natural to consider what happens if one perturbs an extreme RN black hole by dropping something into it. Can one trigger an instability using {\it ingoing} radiation? Some stability results for this case were presented in Ref. \cite{Dain:2012qw}, which considered initial data for $\psi$ compactly supported outside ${\cal H}^+$. It was shown that the scalar field and all of its derivatives remain bounded in {\it static} coordinates. However, such coordinates cover only the black hole exterior so this result does not exclude the existence of an instability at ${\cal H}^+$. 

We have investigated this problem numerically and describe our results in section \ref{Nummassless}. We find that initial data corresponding to an ingoing wavepacket does lead to an instability at ${\cal H}^+$. It afflicts quantities with one more $r$-derivative than in the outgoing case: $\psi$ and $\partial_r \psi$ decay on, and outside ${\cal H}^+$, $\partial_r^2 \psi$ decays outside ${\cal H}^+$ but generically does not decay on ${\cal H}^+$ and $\partial_r^3 \psi$ generically blows up at late time on ${\cal H}^+$. We also study the late time behaviour (tail) of $\psi$ and find that the field decays more slowly if the Aretakis constants are non-zero. This extends previous numerical work on massless scalar field tails in extreme RN \cite{Blaksley:2007ak}, which considered outgoing wavepackets {\it outside} ${\cal H}^+$, which have vanishing Aretakis constants.

It is natural to ask whether the Aretakis instability can be seen in the $AdS_2 \times S^2$ near-horizon geometry of extreme RN. In section \ref{AdS2} we will show that it can: it occurs at the horizon of $AdS_2$ in Poincar\'e coordinates. This is not in contradiction with known stability results for linear fields in $AdS_2$ because it turns out to be a coordinate effect: there is no invariant way of defining coordinates in $AdS_2$ analogous to the $(v,r)$ of (\ref{extremeRN}). Nevertheless, the $AdS_2$ results are interesting because they are in excellent agreement with our numerical results for the late time behaviour of $\psi$ at ${\cal H}^+$ for the extreme RN spacetime.

The remainder of our paper concerns generalisations of Aretakis' work to other fields. In section \ref{Nummassisve} we consider a {\it massive} scalar field. For discrete values of the mass $m$ ($m^2 = n(n+1) M^{-2}$ for positive integer $n$) we prove the existence of conserved quantities analogous to the Aretakis constants and use these to prove instability at ${\cal H}^+$.  For more general values of the mass we demonstrate instability numerically.  A more massive field is more stable: the number of $r$-derivatives exhibiting decay at ${\cal H}^+$ increases with the mass of the field. However, our numerical results indicate that for any mass there exists $k$ such that $\partial_r^k \psi$ generically blows up at late time on ${\cal H}^+$. 

Finally, it is desirable to have an argument for instability of extreme RN that does not rely on the existence of a scalar field. In section \ref{gravem} we will consider (coupled) gravitational and electromagnetic perturbations. We prove that for all types of perturbations of this kind, there exist towers of conserved quantities on the horizon, which can be used to exhibit instabilities very similar to those of the massless scalar field. We construct an explicit gauge invariant combination of the Maxwell field strength perturbation, the metric perturbation and its first derivatives. The instability is strongest for the $l=2$ multipole moment, for which we show that a certain combination of the first and second $r$-derivatives of this quantity is conserved on ${\cal H}^+$ and hence generically cannot decay. We argue that this implies that a quantity with one more $r$-derivative generically blows up at late time on ${\cal H}^+$.  Hence there is an instability at the horizon of extreme RN in Einstein-Maxwell theory (or minimal $N=2$ supergravity). 
 
\bigskip

\noindent{ \bf Note added}: As this paper was nearing completion, Refs.~\cite{Bizon:2012we,newaretakis} appeared. Ref. \cite{Bizon:2012we} has significant overlap with our section \ref{CI} relating the Aretakis and NP constants. Motivated by our numerical results, Ref. \cite{newaretakis} gives a proof of the existence of an instability for ingoing massless scalar field radiation. Ref. \cite{Bizon:2012we} also gives an argument for the existence of such an instability.

\section{Massless scalar horizon instability}

\subsection{Motivation for instability}
\label{redshift}

To understand why there might be an instability at the horizon of an extreme black hole, it is useful to consider first the case of a massless scalar field $\psi$ in a {\it non-extreme} black hole spacetime. A first step in trying to prove stability is to consider the energy of $\psi$, written as an integral $E[\Sigma]$ over a spacelike surface $\Sigma$ extending from the future horizon ${\cal H}^+$ to infinity. This is a non-increasing function of time, i.e., $E[\Sigma'] \le E[\Sigma]$ if $\Sigma'$ lies to the future of $\Sigma$. Outside ${\cal H}^+$, the integrand, i.e., the energy density, is a positive definite function of $\partial_\mu \psi$. Hence the fact that $E$ is non-increasing implies that $\partial_\mu \psi$ cannot become large. However, precisely on ${\cal H}^+$, the energy density degenerates: it does not depend on the derivative of $\psi$ transverse to ${\cal H}^+$. Hence it is consistent with energy conservation for this derivative to behave badly.

For non-extreme black holes, such behaviour has been excluded using the horizon {\it redshift effect}. This is the fact that the energy of a photon moving tangential to ${\cal H}^+$ undergoes a redshift proportional to $e^{-\kappa v}$ where $\kappa$ is the surface gravity and $v$ the Killing time along the horizon. Using the wave analogue of this effect is an important step in the proof that $\psi$ and all of its derivative decay on and outside ${\cal H}^+$ in a Schwarzschild \cite{Dafermos:2008en} or non-extreme Kerr spacetime \cite{Dafermos:2010hd}. 
 
Now consider an extreme black hole. Such a solution has $\kappa=0$: the horizon redshift effect is absent and so the problem of controlling derivatives transverse to ${\cal H}^+$ remains. Aretakis has proved that this problem cannot be overcome for extreme RN and extreme Kerr. Ref. \cite{Lucietti:2012sf} showed that it cannot be overcome for any extreme black hole. Below we will review Aretakis' argument for extreme RN. 

We note that Ref. \cite{bicak} considered the case of a massless scalar field in RN and argued that there is a qualitative difference between the extreme and non-extreme cases arising from the behaviour of outgoing waves near ${\cal H}^+$. However, the detailed predictions of Ref. \cite{bicak} are in disagreement with numerical results of Ref. \cite{Blaksley:2007ak}  and also with the numerical results that we will present below. 

\subsection{Aretakis' argument}

\label{Analytic}

The equation of motion of a massless scalar is
\be
 \nabla^2 \psi = 0 \; .
\ee
Working in the coordinates of (\ref{extremeRN}), first expand $\psi$ in spherical harmonics:
\be
\label{harmonics}
\psi(v,r,\Omega) = \sum_{l=0}^\infty \psi_l (v,r) Y_l(\Omega)
\ee
(we suppress the azimuthal index $m$) and substitute into the equation of motion to obtain
\be
\label{boxpsi}
 2 r \partial_v \partial_r (r\psi_l) + \partial_r \left( \Delta \partial_r \psi_l \right) - l(l+1) \psi_l = 0
\ee
where $\Delta= (r-M)^2$. Set $l=0$ and evaluating at ${\cal H}^+$ ($r=M$) shows that
\be
\label{H0def}
 H_0[\psi] \equiv \frac{1}{M} [\partial_r(r\psi_0)]_{r=M}
\ee 
is conserved, i.e., independent of $v$. 
 
For generic initial data, $H_0$ will be non-zero. Hence it remains non-zero. Therefore $\psi$ and $\partial_r \psi$ cannot both decay at ${\cal H}^+$. In fact, $\psi$ decays at late time \cite{Aretakis:2011ha} hence it follows that its transverse derivative at ${\cal H}^+$ does {\it not} decay:
\be
\label{drpsi}
 (\partial_r \psi_0)_{r=M} \rightarrow H_0 \qquad {\rm as} \qquad v \rightarrow \infty  \; .
\ee
The $rr$ component of the energy momentum tensor of $\psi$ is $T_{rr} = (\partial_r \psi)^2$, which generically does not decay at late time on ${\cal H}^+$. This implies that the energy density measured by an ingoing observer at ${\cal H}^+$ does not decay, as suggested by the absence of the horizon redshift effect. 

Now act on (\ref{boxpsi}) with $\partial_r$, set $l=0$ and evaluate at ${\cal H}^+$ to obtain
\be
 \left[  \partial_v \partial_r^2 (r\psi_0) +  \partial_r \psi_0 \right]_{r=M} = 0 \; .
\ee
Hence
\be
 \left[ \partial_v \partial_r^2 (r\psi_0) \right]_{r=M} \rightarrow - H_0 \qquad {\rm as} \qquad v \rightarrow \infty
\ee
which, together with decay of $\psi_0$ and (\ref{drpsi}), implies blow-up of the second transverse derivative of $\psi$ at late time on ${\cal H}^+$:
\be
( \partial_r^2 \psi_0)_{r=M} \sim -\frac{H_0}{M} v  \qquad {\rm as} \qquad v \rightarrow \infty \; .
\ee
By taking further $r$-derivatives of (\ref{boxpsi}) it can be shown that $( \partial_r^k \psi_0)_{r=M} \propto v^{k-1}$ for large $v$  \cite{Aretakis:2011hc}.

This instability is a property of s-wave ($l=0$) perturbations. There is a corresponding result for $l>0$: acting on (\ref{boxpsi}) with $\partial_r^l$ and evaluating at $r=M$ reveals that
\be
\label{H_l}
 H_l[\psi] \equiv \frac{1}{M^2} \left\{ \partial_r^l \left[ r \partial_r (r \psi_l) \right] \right\}_{r=M}
\ee
is conserved. Aretakis shows that $\partial_r^k \psi_l$ decays on, and outside ${\cal H}^+$ for $k \le l$ \cite{Aretakis:2011ha}. Using this, it follows that $\partial_r^{l+1} \psi_l$ generically does not decay at ${\cal H}^+$ and, arguing as above, $\partial_r^{l+2} \psi_l$ generically blows up linearly at ${\cal H}^+$   \cite{Aretakis:2011hc}. 

\subsection{Conformal isometry}
\label{CI}

The extreme RN solution has a discrete conformal isometry \cite{couch}. In static coordinates $(t,r,\theta,\phi)$ it is
\be
 \Phi: (t,r,\theta,\phi) \rightarrow
 (t,r'=M+\frac{M^2}{r-M},\theta,\phi) 
\label{coniso}
\ee
This is self-inverse: $\Phi = \Phi^{-1}$. The push-forward of the metric under this diffeomorphism is
\be
 \Phi_*(g) = \Omega^2 g
\ee
where
\be
 \Omega= \frac{M}{r-M} \; .
\ee
To understand how this acts on the horizon, we will write it in Eddington-Finkelstein coordinates, related to the static coordinates as follows. First define the tortoise coordinates
\begin{equation}
\label{tortoise}
 r_\ast(r)=r-M+2M\ln\left(\frac{|r-M|}{M}\right) - \frac{M^2}{r-M}\ ,
\end{equation}
and note that $\Phi$ sends $r_*$ to $r_*' = -r_*$. Now let
\be
 u = t-r_*, \qquad v=t+r_*
\ee
and so $\Phi$ sends $u$ to $u'=v$ and $v$ to $v'=u$. Hence $\Phi$ maps the point with ingoing EF coordinates $(v,r,\theta,\phi)$ to the point with {\it outgoing} EF coordinates $(u'=v,r',\theta,\phi)$. It follows that $\Phi$ maps ${\cal H}^+$ to ${\cal I}^+$ and vice-versa.

The extreme RN geometry has vanishing Ricci scalar, which implies that the massless scalar wave equation is conformally covariant: if $\psi$ is a solution in the extreme RN geometry with metric $g$ then $\Omega^{-1} \psi$ is a solution in the geometry with metric $\Omega^2 g$. Hence the conformal isometry can be used to generate a new solution $\tilde{\psi}$ of the massless scalar wave equation from an old one $\tilde{\psi} = \Omega \psi \circ \Phi$. In coordinates:
\be
 \tilde{\psi}(u,r,\theta,\phi)_O = \frac{M}{r-M} \psi(v'=u,r',\theta,\phi)_I
\ee
On the left, the subscript $O$ indicates that outgoing EF coordinates are used, on the right the subscript $I$ indicates that ingoing EF coordinates are used.

What happens to the Aretakis conserved quantities under this map? Long ago Newman and Penrose (NP) argued that there is an infinite set of conserved quantities associated to linear massless fields at ${\cal I}^+$ \cite{Newman:1968uj}. These can be defined by developing an asymptotic expansion in inverse powers of $r$ near ${\cal I}^+$: for the $l=0$ multipole moment of $\psi$ assume the expansion
\be
 \psi_0 \sim  \frac{f_{0}(u)}{r} + \frac{f_{1}(u)}{r^{2}} 
\label{NPdef}
\ee
as $r \to \infty$.\footnote{This can be deduced if we assume that $\Omega^{-1} \psi$ is smooth at ${\cal I}^+$ in the conformally compactified spacetime.}  The equation of motion for $\psi$ implies that the quantity $f_{1}$ is independent of $u$: this is the NP constant of $\psi_0$ which we will denote by $f_{1}[\psi]$. It is easy to check that the conformal isometry exchanges the Aretakis conserved quantity with the NP constant, i.e., the Aretakis conserved quantity for $\psi_0$ is the NP constant of $\tilde{\psi}_0$ and vice versa. Explicitly we find
\be
H_0[\psi] = \frac{f_{1}[\tilde{\psi}] }{M^3}  \; .
\ee
The physical interpretation of the NP constants is not well-understood but it is clear that they are closely related to {\it ingoing} radiation at ${\cal I}^+$, just as the Aretakis constants are related to outgoing radiation at ${\cal H}^+$. The NP constants influence the decay rate (tail) of $\psi$ outside the horizon: there is evidence that initial data with non-vanishing NP constants results in slower decay than data with vanishing NP constants \cite{Gomez:1994ns}. We will return to this point below.

As explained above, if any of the Aretakis conserved quantities of $\psi$ is non-zero then $\psi$ has an instability at ${\cal H}^+$. Does this mean that there is an instability of $\psi$ at ${\cal I}^+$ if its NP constants are non-zero? Even in Minkowski spacetime, there is a late-time blow-up of transverse derivative of $\Omega^{-1} \psi$ at ${\cal I}^+$ when the NP constants are non-zero \cite{Newman:1968uj}. However, unlike ${\cal H}^+$, ${\cal I}^+$ is not part of the physical spacetime hence this is not an instability: the field and all its derivatives decay in the physical spacetime. 

The definition of the NP constants has been criticized because of the assumed smoothness of $\Omega^{-1} \psi$ at ${\cal I}^+$ \cite{Bardeen:1973xb}. For spacetimes with non-vanishing NP constants, generically one expects only a finite degree of differentiability at ${\cal I}^+$. In contrast, $\psi$ will be smooth at ${\cal H}^+$ if it arises from smooth initial data prescribed on a surface $\Sigma$ which extends a finite distance behind ${\cal H}^+$ as in Fig.~\ref{penrose}. Thus there is an asymmetry present in the amount of differentiability to be expected at ${\cal H}^+$ and ${\cal I}^+$. Starting from initial data specified on $\Sigma$ we get a solution $\psi$ smooth at ${\cal H}^+$ but generically not smooth at ${\cal I}^+$. Applying the above map then gives a solution $\tilde{\psi}$ for which $\Omega^{-1} \tilde{\psi}$ is smooth at ${\cal I}^+$ but $\tilde{\psi}$ is not smooth at ${\cal H}^+$. Such $\tilde{\psi}$ does not correspond to smooth initial data on $\Sigma$ and therefore lies outside the class of solutions considered in this paper.

\section{Numerical results for massless scalar}
\label{Nummassless}

\subsection{Double null coordinates}

A crucial assumption in Aretakis' proof of instability is that
the conserved quantities $H_l[\psi]$ are non-vanishing. This
requires that the scalar field $\psi$ be non-vanishing at the intersection of ${\cal H}^+$ and the hypersurface $\Sigma$  on which initial data is specified. Hence the results in section \ref{Analytic} do not reveal what happens if one
perturbs an extreme RN black hole by sending waves into the black hole
from outside. We will investigate this problem numerically. 
To do this we first introduce coordinates which are better suited to numerical evolution.

In $(u,v)$-coordinates the  extreme RN black hole metric is
\begin{equation}
 ds^2=-F(r(u,v))dudv+r(u,v)^2 d\Omega^2\ ,
\end{equation}
where
\be
 F(r) = \left( 1 - \frac{M}{r} \right)^2
\ee
and $r(u,v)$ can be determined by solving $r_\ast(r)=(v-u)/2$ with $r_\ast$ defined by (\ref{tortoise}). 

This metric is singular at ${\cal H}^+$. Since we wish to investigate time evolution of perturbations on the
horizon, we need a regular metric there.
Thus, we introduce a further coordinate transformation defined by
\begin{equation}
 \frac{u}{2}=-r_\ast(M-U)=U-2M\ln\left(\frac{|U|}{M}\right) - \frac{M^2}{U}\ .
\end{equation}
From the definition of $r_*$, we have $du/dU=2/F(M-U)$.
In $(U,v)$-coordinates $\mathcal{H}^+$ is located at
$U=0$ and and $U<0$ corresponds to the exterior of the black hole. In this coordinate system the metric reads
\begin{equation}
 ds^2=-\frac{2F(r)}{F(M-U)}dUdv + r^2d\Omega^2\ ,
\label{regmet}
\end{equation}
where $r=r(U,v)$. 
Using $r_\ast(r)=(v-u)/2$ it can be seen that $r$ is analytic in $U,v$. In particular we can expand $r$ for small $U$ (and fixed $v$) to obtain
\begin{equation}
 r-M=-U+\frac{v}{2M^2}U^2+\left(\frac{v}{M^3}-\frac{v^2}{4M^4}\right)U^3+\cdots\ .
\end{equation}
It follows that $F(r)/F(M-U)= 1 +\mathcal{O}(U)$ for small $U$ and the metric~(\ref{regmet}) is analytic at $\mathcal{H}^+$, i.e., it can be analytically continued to the black hole interior $U>0$. 

\subsection{Klein-Gordon equation and initial data}

Consider the Klein-Gordon equation
\be
\label{KG}
(\nabla^2-m^2)\psi=0
\ee
in the extreme RN background.
Defining $\phi\equiv r\psi_l$, where $\psi_l$ is the $l$th multipole moment of $\psi$ as in equation (\ref{harmonics}), we obtain a wave equation for $\phi$. In  the $(U,v)$-coordinates introduced above this reads
\bea
\label{waveUv}
 -4\partial_U \partial_v \phi  &=&  \hat V(U,v)\phi \\ 
 \hat
 V(U,v) &=& \frac{2F(r)}{F(M-U)}\left(\frac{F'(r)}{r}+\frac{l(l+1)}{r^2}+m^2\right)\ .  \nonumber
\eea
In this section, we study the massless scalar field $m=0$.
We will consider the massive scalar in section \ref{Nummassisve}.

\begin{figure}
\begin{center}
\includegraphics[scale=0.4]{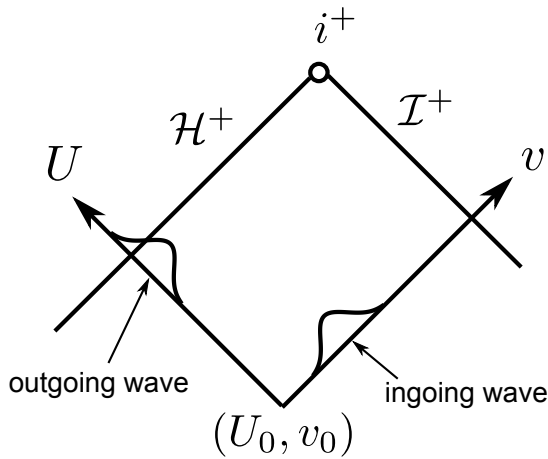}
\end{center}
\caption{
Schematic plot of the setup used in the numerical calculations.
}
 \label{initial}
\end{figure}

We consider a null ``initial''
surface defined by
\be
\Sigma_0 = \{ U=U_0, v \geq v_0 \} \cup \{   U \geq U_0,  v=v_0 \}
\ee
and impose the following initial data:
\begin{equation}
\label{out}
 \phi(U,v_0)=\exp\left(-
\frac{(U-\mu)^2}{2\sigma^2}
\right)\ ,  
\qquad
\phi(U_0,v)=0\ ,
\end{equation}
or 
\begin{equation}
\label{ing}
 \phi(U,v_0)=0\ ,  
\qquad
\phi(U_0,v)=\exp\left(-
\frac{(v-\mu')^2}{2\sigma'{}^2}
\right)\ ,
\end{equation}
which correspond to an outgoing and ingoing wavepacket, respectively.  In Fig.~\ref{initial} we give a schematic plot of the setup.

We will solve this problem numerically.  In our numerical calculations, we use units
such that $M=1$ and set $U_0=-0.5$ and $v_0=0$.
We describe the details of this in Appendix \ref{ALG}.

\subsection{Numerical simulation for spherically symmetric mode}

In this section we study spherically symmetric solutions, i.e., we set $l=0$. In this case Aretakis' conserved quantity defined in Eq.~(\ref{H0def}) is simply given by  $H_0[\psi]=\partial_r\phi|_{r=M}$.  Hence the outgoing wave initial data (\ref{out}) has $H_0[\psi] \neq0$ (unless $\mu=0$) whereas the ingoing wave initial data (\ref{ing}) has $H_0[\psi]=0$ (since its support does not intersect the horizon).

\subsubsection{Non-zero Aretakis constant}
\label{Sec:massless_nonzero}

Firstly, we consider solutions for which $H_0[\psi]\neq 0$.
Although, in this case, 
the instability of the scalar field has been shown
analytically, as reviewed in section \ref{Analytic},
we will investigate finer details of the time evolution of the scalar field by
solving the wave equation numerically. 

Although we used the $(U,v)$ coordinates for our numerical calculations, we will display our results using $(v,r)$ coordinates. The reason for this is that the $(v,r)$ coordinates are a preferred set of coordinates associated to the symmetries of the background. In particular, $\partial/\partial v$ is the generator of time translations and $-\partial/\partial r$ is tangent to ingoing radial null geodesics of unit energy. The $(U,v)$ coordinates are not so closely associated to geometric invariants e.g.\ the generator of time translations is not $\partial/\partial v$ in these coordinates. Hence when working in $(U,v)$ coordinates it is less clear whether something is a physical effect or merely a coordinate effect.

\begin{figure}
  \centering
  \subfigure
  {\includegraphics[scale=0.5]{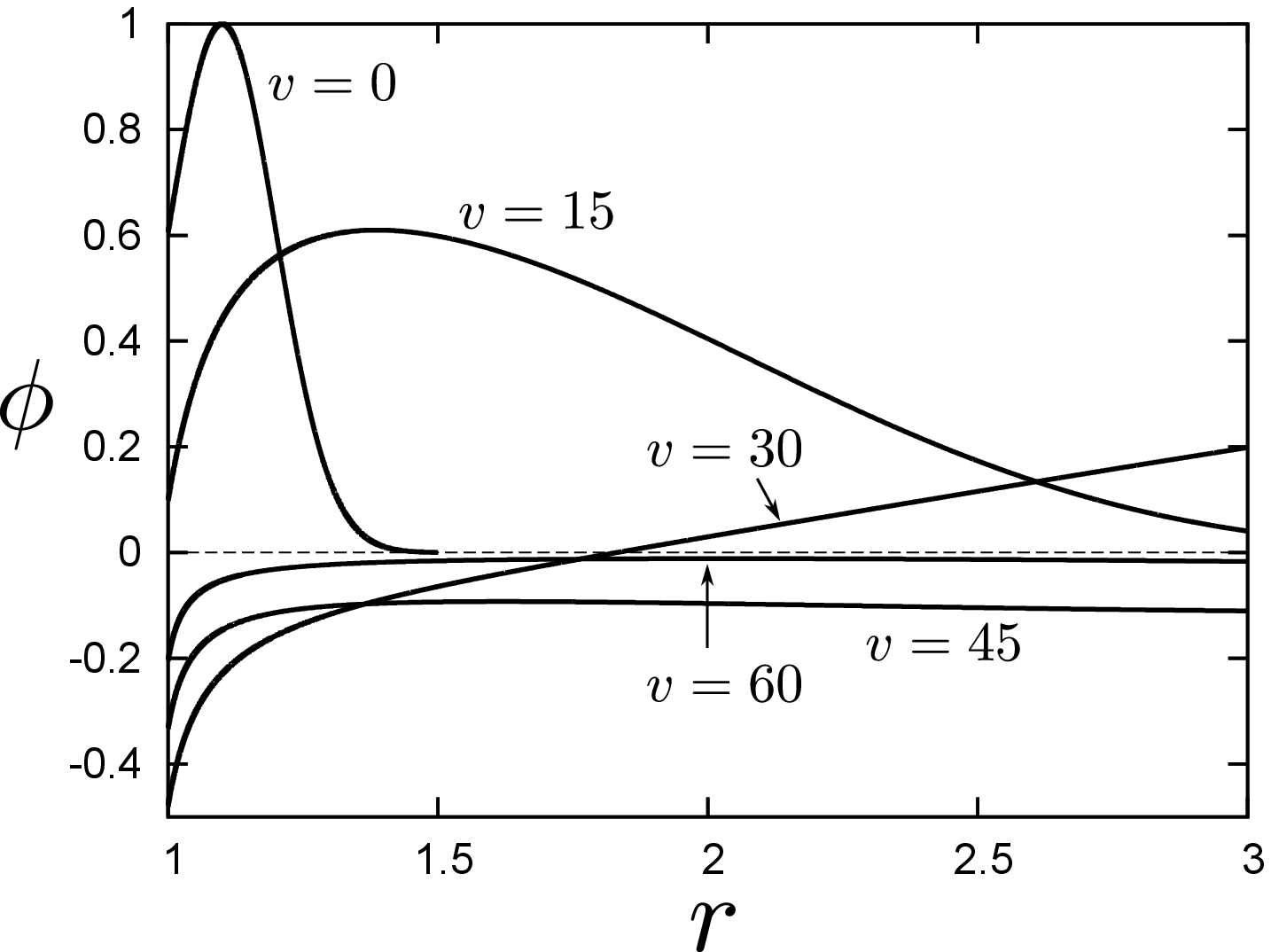}
\label{phi_ev1}
  }
  \subfigure
  {\includegraphics[scale=0.5]{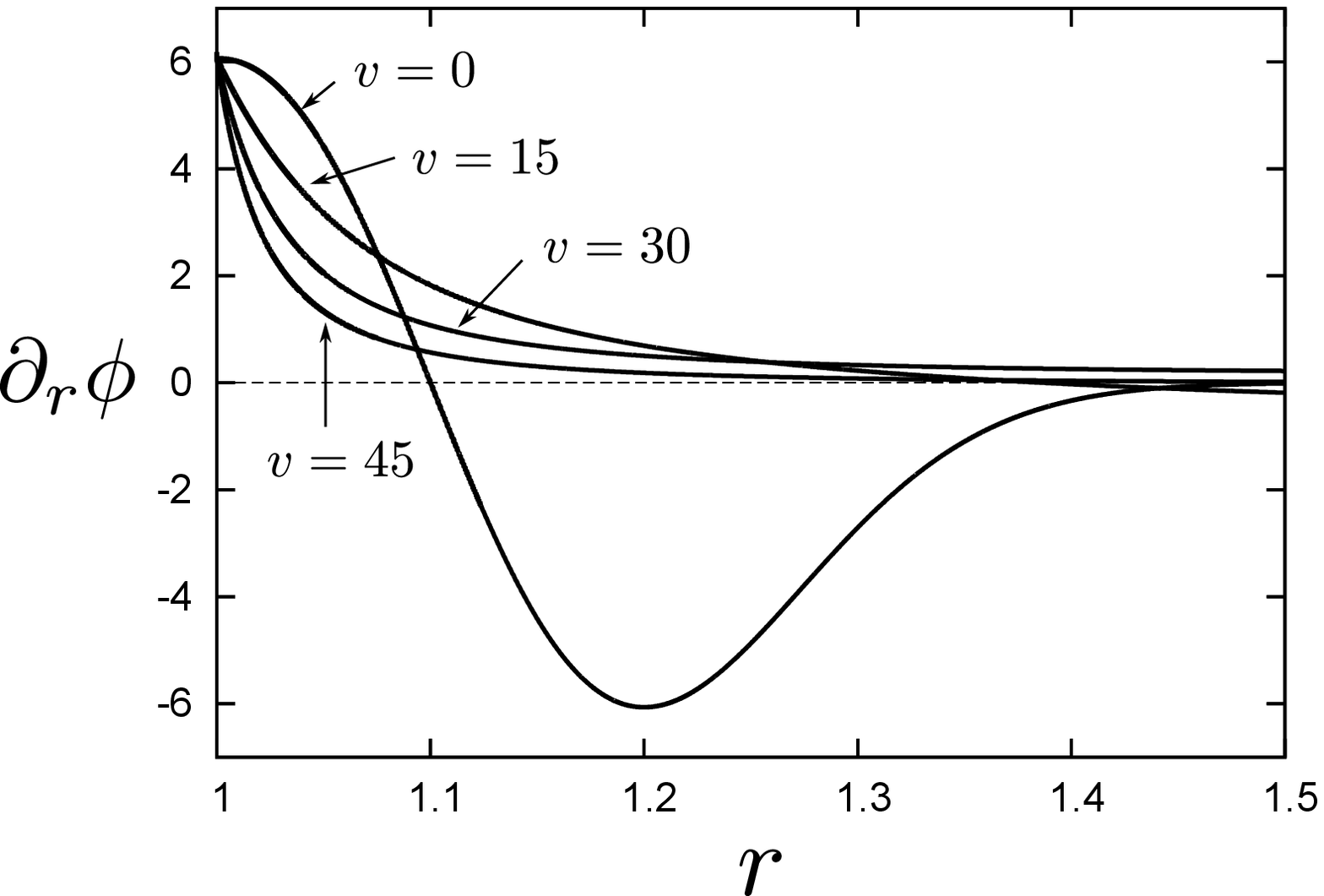} 
\label{phi_ev2}
  }
  \caption{
Functions 
$\phi(v,r)$ and $\partial_r\phi(v,r)$
for $l=0$ and $H_0\neq0$ 
on fixed $v$ slices.
We consider outgoing wave initial data~(\ref{out}) with
$(\sigma,\mu)=(0.1,-0.1)$. The horizon is at $r=1$.
We can see that $\partial_r\phi$ becomes steeper near the horizon
as $v$ increases. This implies that $\partial_r^2\phi$ blows up on the
horizon at large $v$.
\label{phi_ev}
}
\end{figure}

Consider outgoing initial data~(\ref{out}) with
$(\sigma,\mu)=(0.1,-0.1)$. The time evolution of $\phi$ and
$\partial_r\phi$ in $(v,r)$-coordinate is plotted in 
Fig.~\ref{phi_ev}.%
We can see that $\phi$ decays as $v$ increases. On the other hand,
$\partial_r\phi$ does not decay along the horizon because $\partial_r\phi|_{r=M}$ must be conserved. However, 
outside the horizon,
$\partial_r\phi$ decays.
As a result, $\partial_r\phi$ becomes steeper near the horizon 
as time increases.
This is consistent with the fact that $\partial_r^2\phi|_{r=M}$ must blow up along the horizon.

\begin{figure}
  \centering
  \subfigure
  {\includegraphics[scale=0.45]{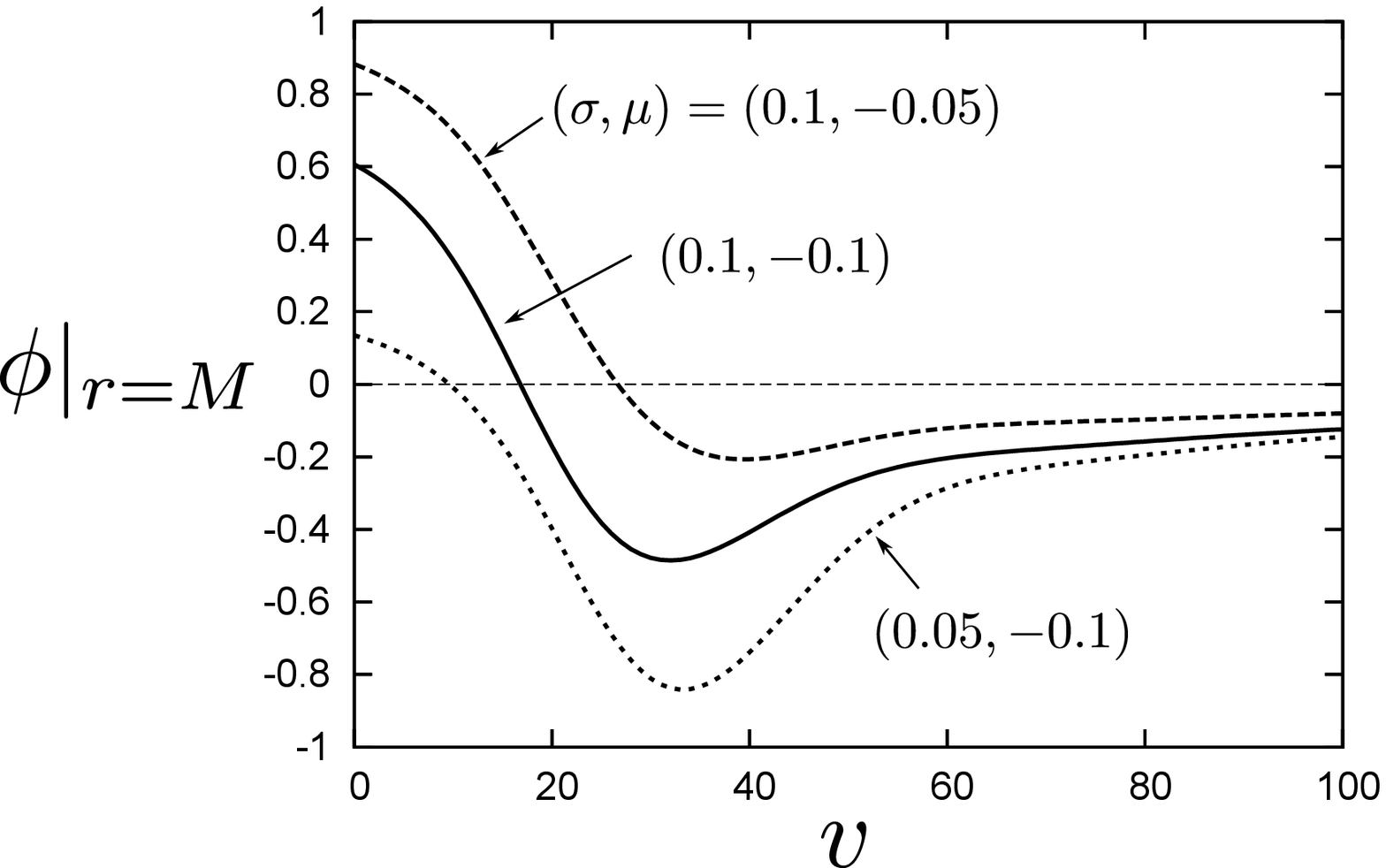}
  }
  \subfigure
  {\includegraphics[scale=0.45]{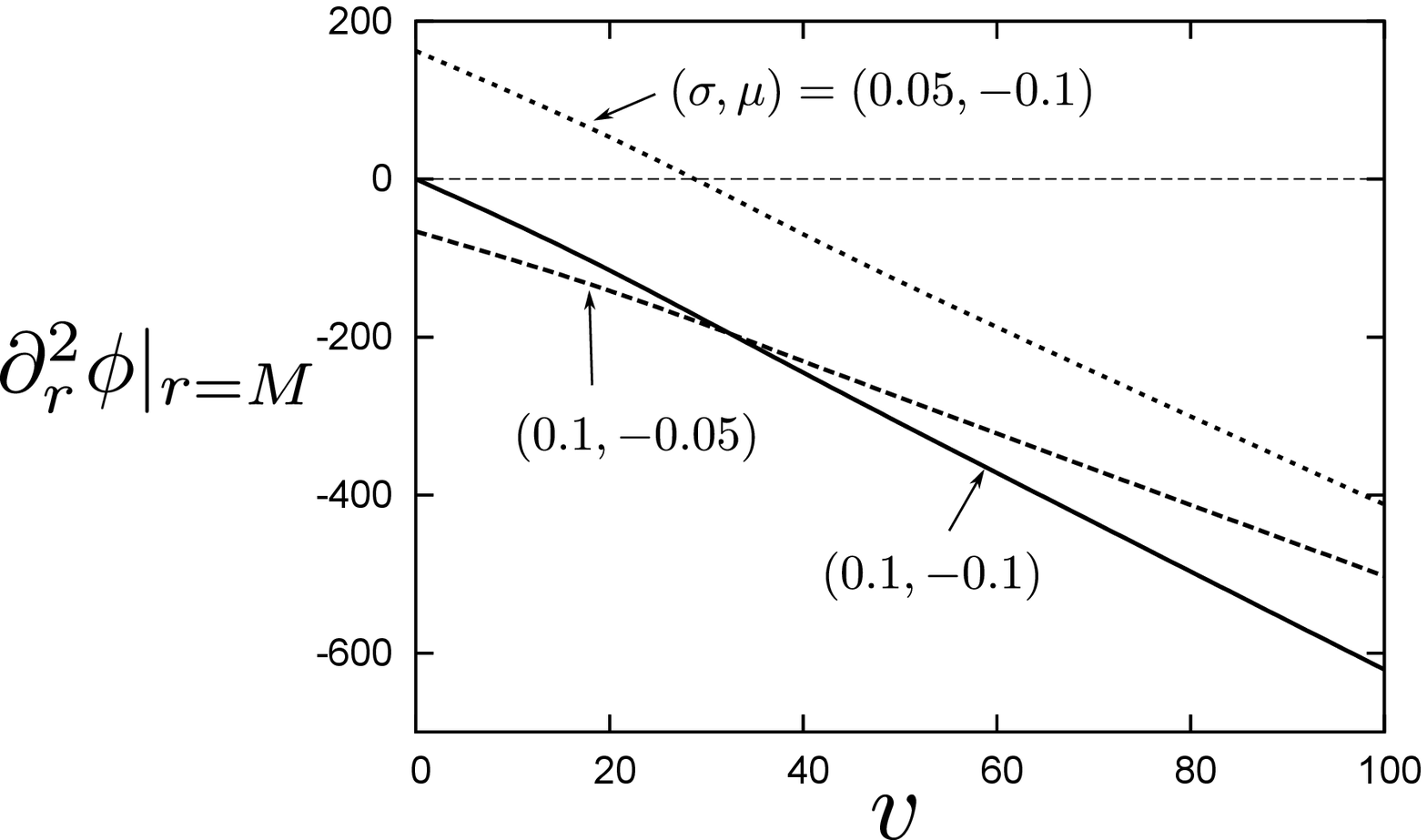} 
  }
  \caption{
Time dependence of $\phi|_{r=M}$ and $\partial_r^2\phi|_{r=M}$ 
for $l=0$ and $H_0\neq 0$. We choose parameters as
 $(\sigma,\mu)=(0.1,-0.1),(0.05,-0.1),(0.1,-0.05)$ 
in the outgoing wave initial data~(\ref{out}).
We can see that $\partial_r^2\phi|_{r=M}$ blows up as $\sim v$.
\label{phiH_ev}
}
\end{figure}

In Fig.~\ref{phiH_ev}, we plot the time evolution of $\phi|_{r=M}$ and
$\partial_r^2\phi|_{r=M}$ for various initial data:
$(\sigma,\mu)=(0.1,-0.1),(0.05,-0.1),(0.1,-0.05)$. 
We can see that the time behaviour of 
$\partial_r^2\phi|_{r=M}$ is consistent with the expected linear blow up shown in section~\ref{Analytic}.
In Appendix \ref{ETD}, we describe our method for evaluating the transverse
derivatives on the horizon $\partial_r^n \phi|_{r=M}$.

Now, we investigate the late time behaviour of $\phi|_{r=M}$.
By fitting the absolute value of $\phi|_{r=M}$ to a power law decay $v^{a}$, 
we obtain the following exponents:
$a=-0.999, -1.018, -0.991$
for $(\sigma,\mu)=(0.1,-0.1),(0.05,-0.1),(0.1,-0.05)$, 
respectively.\footnote{
The exponents do not change up to the third decimal place 
even if we change the grid size as 
$(\delta U, \delta v)\to (4\delta U, 5\delta v)$.
Thus, the numerical error mainly comes from truncations.
The truncation error can be estimated as $\mathcal{O}(1/v)\sim 10^{-3}$.
The exponents obtained here are reliable up to around second decimal
place.
}
We used the numerical data in the range $1000\leq v \leq 2000$ for the fitting.
These results suggest that the scalar field on the horizon $\phi|_{r=M}$ decays as
$v^{-1}$ at late time.
To determine the coefficient of the power law decay,
we fit the $\phi|_{r=M}$ to the function $\alpha H_0/v$ for $1000\leq v \leq 2000$.
Then, we find the fitting parameter
$\alpha=-1.997,-2.035,-1.979$ for
$(\sigma,\mu)=(0.1,-0.1),(0.05,-0.1),(0.1,-0.05)$, respectively.
Hence, for all cases that we have considered, the leading term
of the late time behaviour of $\phi|_{r=M}$ is given by
\begin{equation}
 \phi|_{r=M} \sim -\frac{2.0 H_0}{v} \qquad v\to \infty \ .
\label{latephi}
\end{equation}
In Sec.~\ref{AdS2} we will show that the same expression can be obtained analytically from an $AdS_2$
calculation.
In Ref. \cite{Aretakis:2011hc} 
it was shown analytically 
that the scalar field on the horizon at late time decays at least as fast as $v^{-3/5}$, which is consistent with our results.
All of our initial data has vanishing NP constant. We will explain below why the coefficient of $v^{-1}$ must be modified for initial data with non-vanishing NP constant.

We can relate this late time behaviour to earlier results of Ref. \cite{Gomez:1994ns} by using the conformal isometry~(\ref{coniso}).\footnote{
We are grateful to the referee for noticing the connection between our results and those of Ref. \cite{Gomez:1994ns}.} Applying this map to our solutions gives solutions which are smooth\footnote{
 See the end of section \ref{CI} for a discussion of why smoothness at ${\cal I}^+$ is non-trivial.}  at ${\cal I}^+$ whose late time behaviour at ${\cal I}^+$ is determined by our result for the late time behaviour at ${\cal H}^+$. In particular, the result $\phi \sim -2H_0/v$ at ${\cal H}^+$ maps to the result $\phi\sim -2f_1/u$ at ${\cal I}^+$ where $f_1$ is the NP constant defined in Eq.~(\ref{NPdef}). This agrees with the late time behaviour found in Ref. \cite{Gomez:1994ns}, which studied a massless scalar field in the region $r>R$ of the {\it Schwarzschild} spacetime, with reflecting boundary conditions imposed on the surface $r=R>2M$. For small $M/R$, it was shown perturbatively that
\be
\label{phiscriplus}
 \phi\sim -\frac{2f_1}{u}+\frac{4 M f_1\ln u}{u^2}+\mathcal{O}(u^{-2}) \qquad (M/R \ll 1)
\ee
This perturbative result was confirmed by solving numerically for the scalar field. 

\begin{figure}
\begin{center}
\includegraphics[scale=0.4]{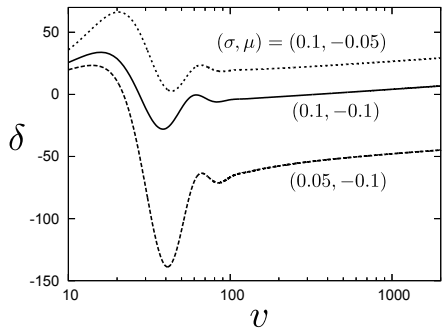}
\end{center}
\caption{
Time dependence of $\delta$. We can see that $\delta$ is linear in
 $\log v$ at late time.
}
 \label{logterm}
\end{figure}

The result (\ref{phiscriplus}) suggests that, in our case, we should find $\ln v/v^2$ corrections to (\ref{latephi}) at next order in $v$. To investigate this, define 
\begin{equation}
 \delta=\frac{v^2}{H_0}\left(\phi|_{r=M}+\frac{2H_0}{v}\right)\ .
\end{equation}
We plot $\delta$ as a function of $v$ in Fig.~\ref{logterm}. This suggests that $\delta$ is a linear function of 
$\ln v$ at late time. 
Thus, we have (with $M=1$)
\begin{equation}
 \phi|_{r=M} \sim -\frac{2.0 H_0}{v} + \frac{\gamma H_0\ln v}{v^2}\qquad v\to \infty \ .
\label{latephi_second}
\end{equation}
To determine the coefficient of the logarithmic term $\gamma$, 
we fit $\delta$ to the function 
$\gamma \ln v + \gamma'$ for $1000\leq v \leq 2000$ and obtain
$\gamma=3.88$, $4.35$, $3.89$
for $(\sigma,\mu)=(0.1,-0.1)$, $(0.05,-0.1)$, $(0.1,-0.05)$,
respectively. These numbers are probably accurate only to 1 significant figure so $\gamma=4$. This is consistent with the expectation that $\gamma$ is independent of the initial data (because if one subtracts two solutions with the same $H_0$ one obtains a solution with vanishing $H_0$ for which we show below that $\phi|_{r=M} \sim v^{-2}$).\footnote{Similarly $\Gamma$ must be independent of the initial data because if one subtracts two solutions with the same $f_1$ one obtains a solution with vanishing $f_1$, for which $\phi \sim u^{-2}$ at ${\cal I}^+$ \cite{Gomez:1994ns}.} After applying the conformal isometry we obtain $\phi\sim -2f_1/u + \gamma f_1 \ln u/u^2$ at ${\cal I}^+$ which has the same form as (\ref{phiscriplus}) ($M=1$ here). Furthermore, our result $\gamma=4$ is in striking agreement with (\ref{phiscriplus}). 

In summary, using the conformal isometry, our results for the late time behaviour at ${\cal H}^+$ can be mapped to late time behaviour at ${\cal I}^+$ of solutions smooth at ${\cal I}^+$. Our results for extreme RN agree with the perturbative result (\ref{phiscriplus}) for Schwarzschild with a reflecting boundary. This agreement suggests that the behaviour (\ref{phiscriplus}) is universal in the sense that it will hold for a larger family of asymptotically flat spacetimes, at least for fields which are smooth at ${\cal I}^+$. 

\subsubsection{Zero Aretakis constant}

\begin{figure}
  \centering
  \subfigure
  {\includegraphics[scale=0.45]{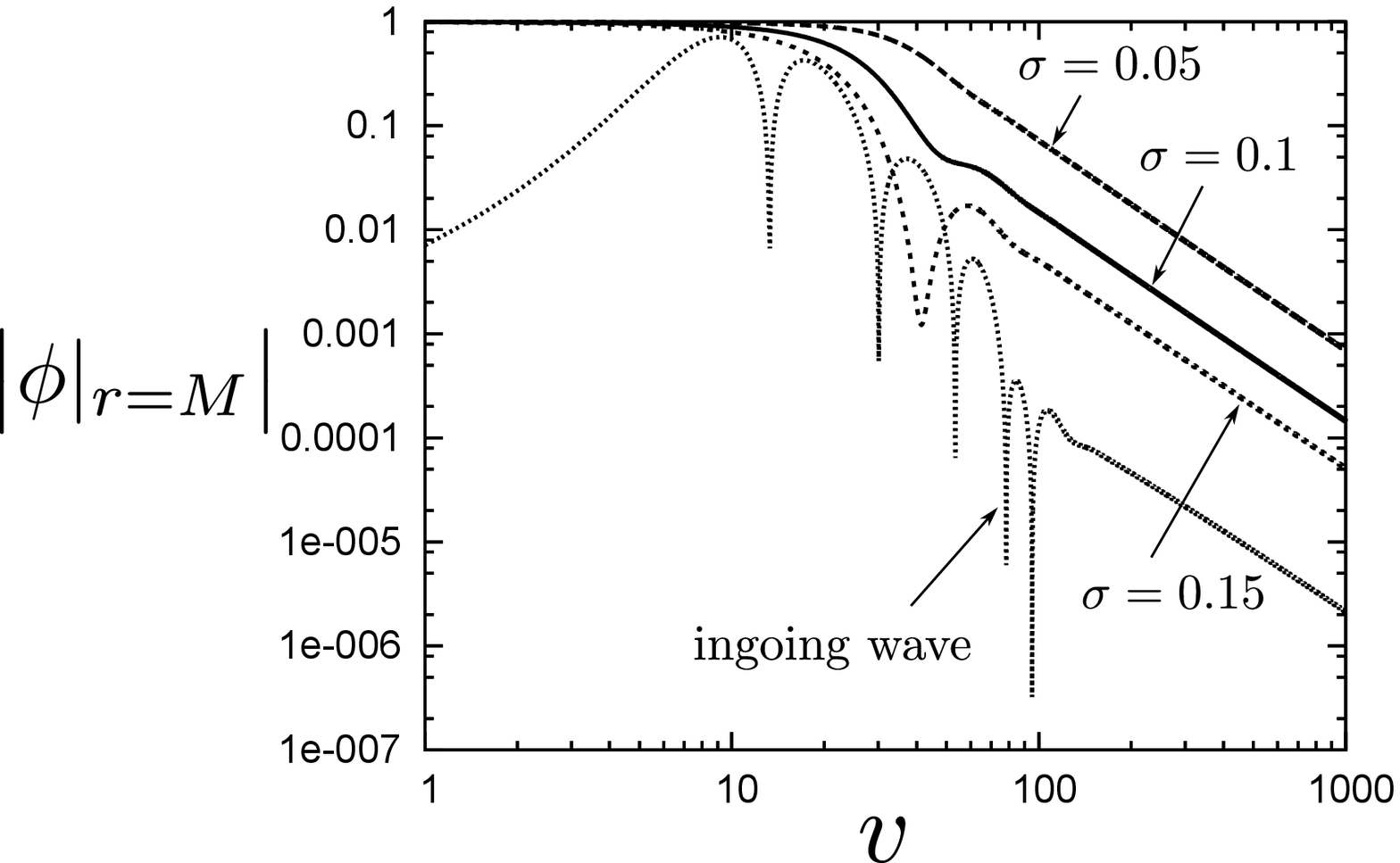}
  }
  \subfigure
  {\includegraphics[scale=0.45]{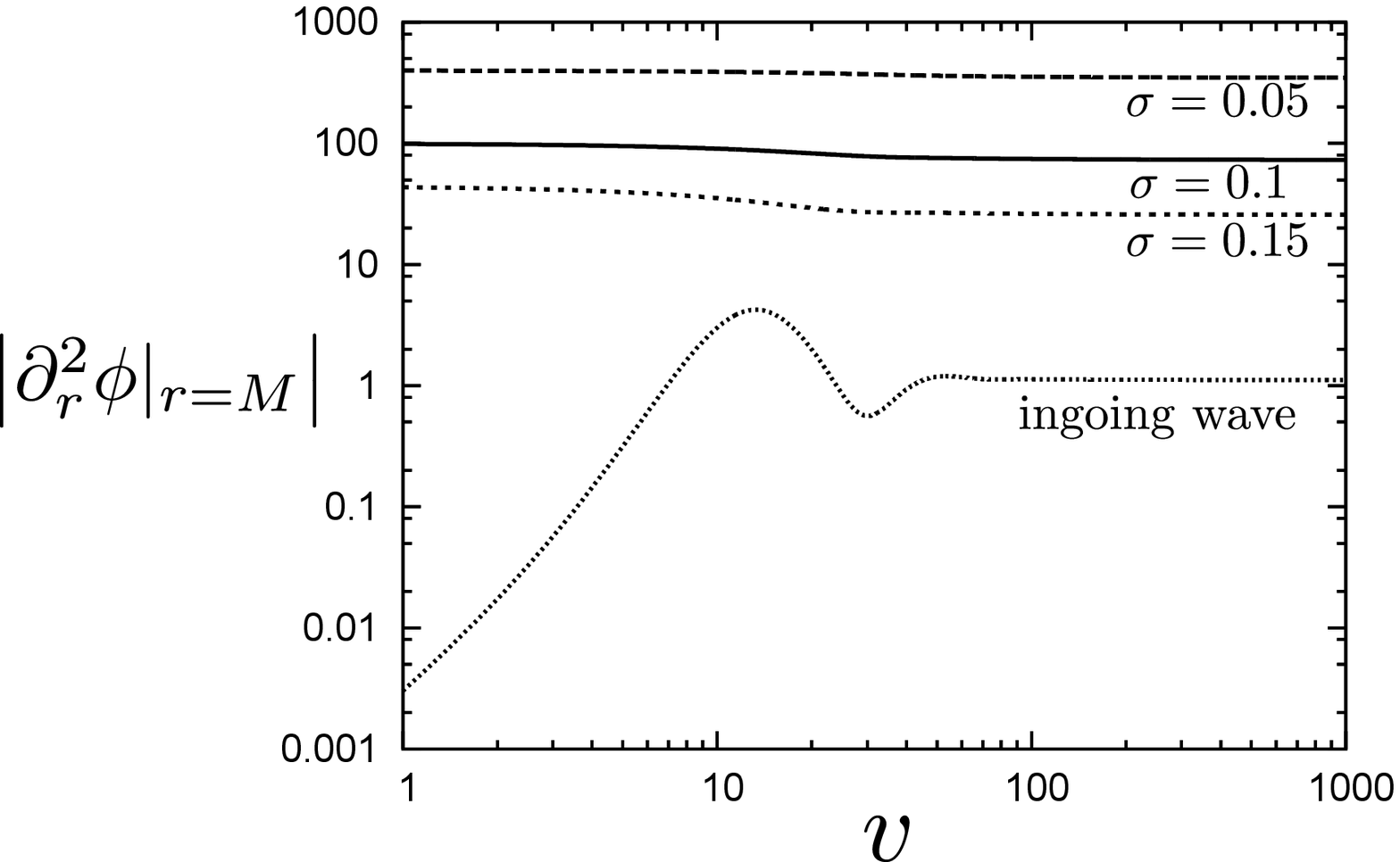} 
  }
  \subfigure
  {\includegraphics[scale=0.45]{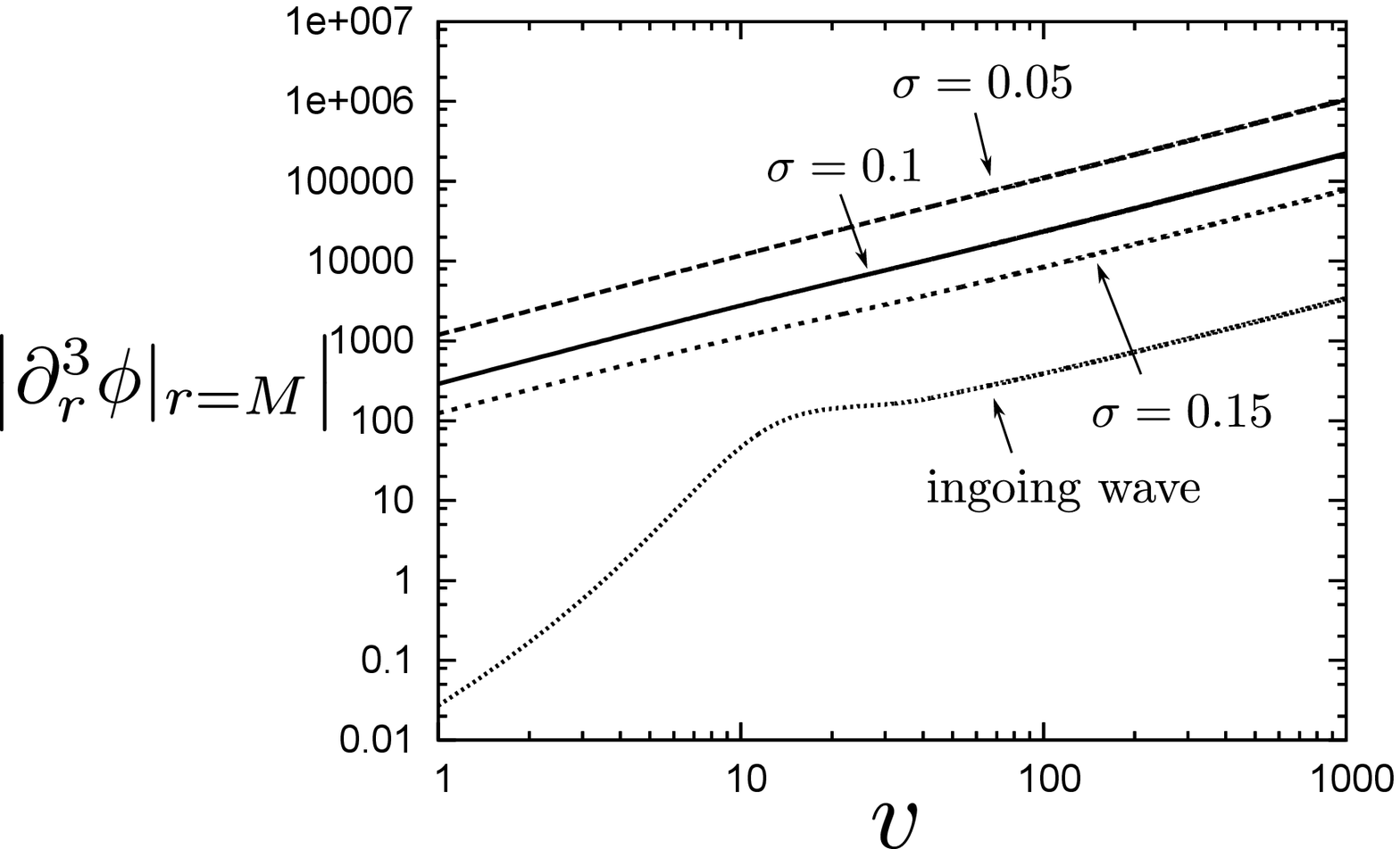} 
\label{phiH_ev3}
  }
\caption{
We depict $\phi|_{r=M}$, $\partial_r^2 \phi|_{r=M}$ and $\partial_r^3 \phi|_{r=M}$ 
as functions of $v$ 
for $l=0$ and $H_0=0$.
We take several initial data: 
$\sigma=0.05,0.1,0.15$ and $\mu=0$ for an outgoing wavepacket and 
$(\sigma',\mu')=(3.0,10.0)$ for an ingoing wavepacket. For the ingoing wavepacket there is a period of damped oscillation (quasinormal ringing) before the power-law tail behaviour takes over.  
In all cases we find that, for large $v$, $\phi|_{r=M}$ decays as $v^{-2}$, while
$\partial_r^2 \phi|_{r=M}$ approaches a non-zero constant and 
$\partial_r^3 \phi|_{r=M}$ blows up as $v$. 
This implies that there is an instability even 
for initial data with $H_0[\psi]=0$.
}
 \label{phiHl0I0}
\end{figure}

Now consider the case where $H_0[\psi]=0$. This includes the case of an ingoing wavepacket as well as an outgoing wavepacket with $\mu=0$. The proof of the Aretakis instability in section \ref{Analytic} does not apply now; instead we will use a numerical calculation to study this case.
For the outgoing wave~(\ref{out}), we take $\mu=0$ and $\sigma=0.05,0.1,0.15$.
For the ingoing wave~(\ref{ing}), we choose the parameters 
$(\sigma',\mu')=(3.0,10.0)$.
In Fig.~\ref{phiHl0I0} 
we plot $\phi|_{r=M}$, $\partial_r^2 \phi|_{r=M}$
and $\partial_r^3 \phi|_{r=M}$ as functions of $v$.
Our results are consistent with the following behaviour: $\phi|_{r=M}$ decays, 
$\partial_r^2 \phi|_{r=M}$ approaches a non-zero constant and 
$\partial_r^3 \phi|_{r=M}$ blows up proportionally to $v$. 
Thus, we conclude that there is an instability even 
for initial data with $H_0[\psi]=0$.

Now consider the late time behaviour of $\phi|_{r=M}$.
Fig.~\ref{phiHl0I0} suggests that $\phi|_{r=M}$ decays as a power law $v^{a}$ at late time.
Fitting to such a decay law we find the exponents
$a=-2.000$, $-1.997$, $-1.995$ for the outgoing wave with $\sigma=0.05,0.1,0.15$  and $a=-1.975$ for the ingoing wave.
This suggests that, for $H_0=0$, the late
time behaviour of $\phi|_{r=M}$ is given by
\begin{equation}
 \phi|_{r=M}\sim \frac{C}{v^2} \qquad
v\to\infty\ .
\end{equation}
We could not find a simple expression for $C$: it may depend on the initial
data in a complicated way. 

\subsection{Numerical simulation for $l=1$}

\subsubsection{Non-zero Aretakis constant}

\begin{figure}
  \centering
  \subfigure
  {\includegraphics[scale=0.45]{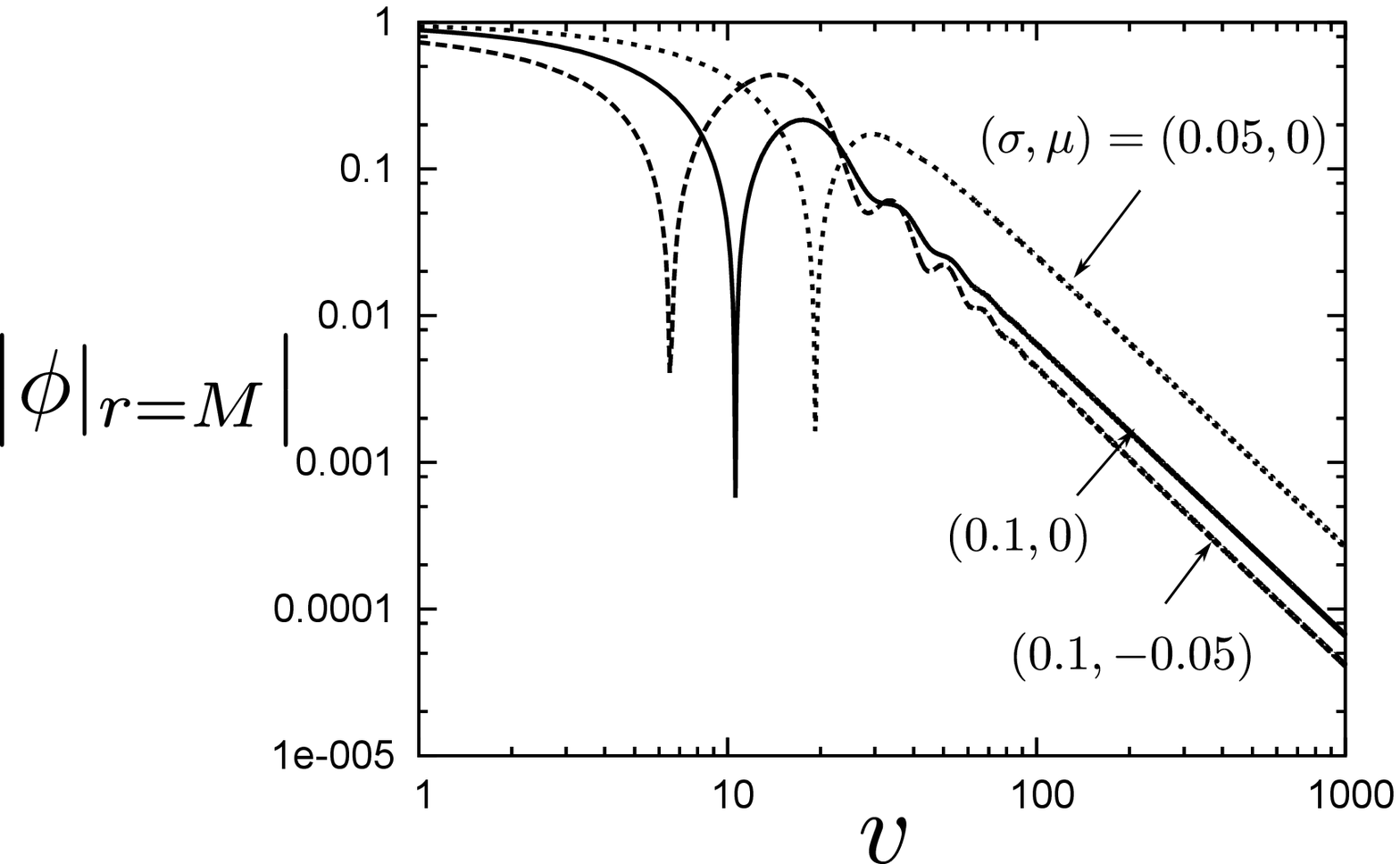}
\label{phiH_l1}
  }
  \subfigure
  {\includegraphics[scale=0.45]{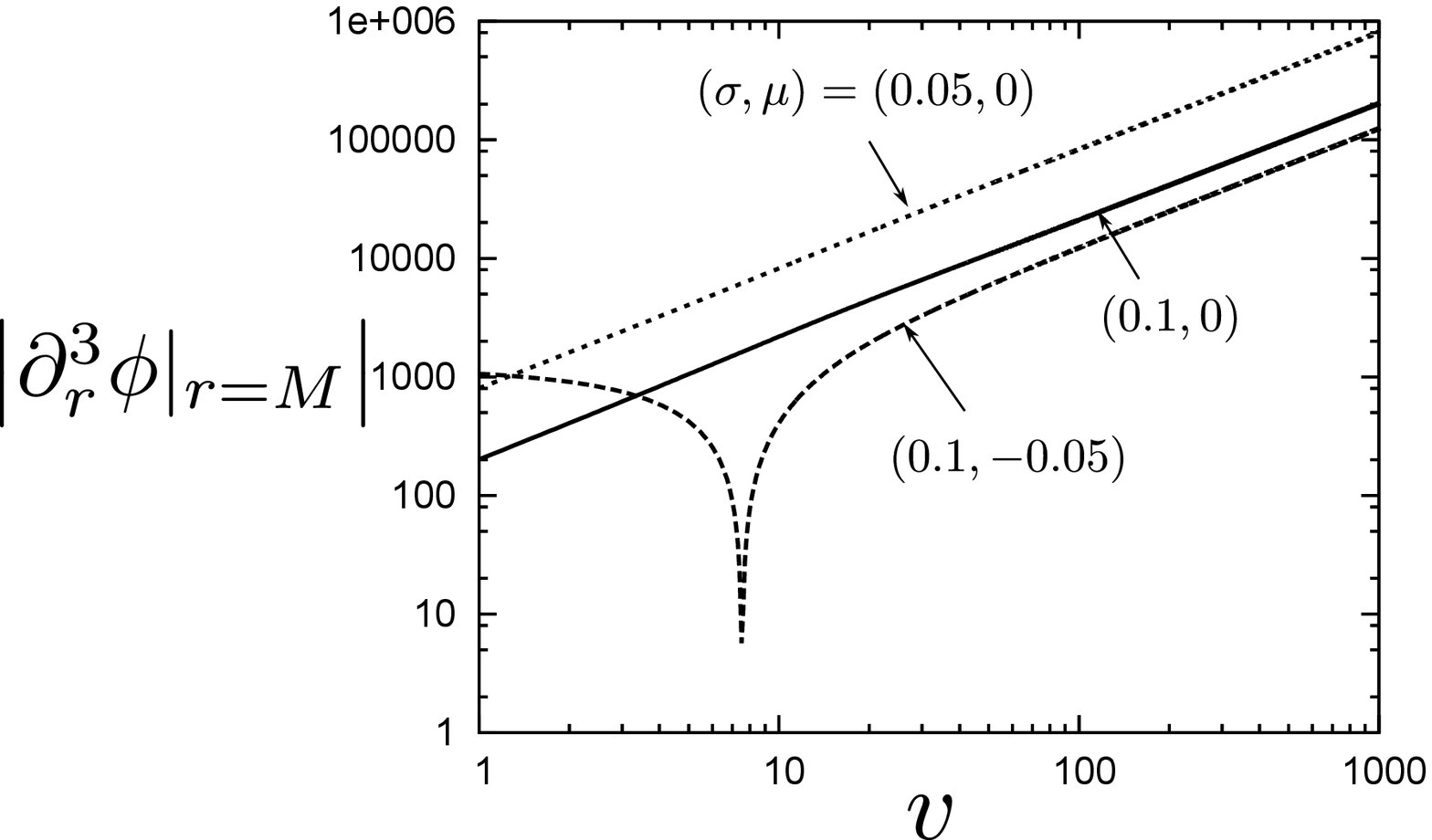} 
\label{phirrrH_l1}
  }
  \caption{
Time dependence of $\phi|_{r=M}$ and $\partial_r^3\phi|_{r=M}$ 
for $l=1$ and $H_1\neq 0$.
We consider outgoing wavepackets with $(\sigma,\mu)=(0.1,0),(0.1,-0.05),(0.05,0)$.
After some damped oscillations (quasinormal ringing), $\phi|_{r=M}$ exhibits power-law decay while $\partial_r^3 \phi$ 
blows up linearly in $v$.
\label{phiH_l1_all}
}
\end{figure}

Now we consider the case of the $l=1$ partial wave, first with $H_1 \neq 0$.
For the outgoing wave we choose the parameters in the initial data to be
$(\sigma,\mu)=(0.1,0),(0.1,-0.05),(0.05,0)$.
In Fig.~\ref{phiH_l1_all} we plot $\phi|_{r=M}$ and 
$\partial_r^3 \phi|_{r=M}$
as functions of $v$. 
We can see that $\phi|_{r=M}$ decays and $\partial_r^3 \phi|_{r=M}$ 
blows up linearly in $v$, as proved by Aretakis.

Next we consider the late time behaviour of $\phi|_{r=M}$.
Fig.~\ref{phiH_l1_all} is consistent with a power law decay of
$\phi|_{r=M}$ of the form $v^{a}$ for large $v$.
Fitting the absolute value of $\phi|_{r=M}$ to such a power law in the range $1000\leq v \leq 2000$,
we obtain the exponents
$a=-1.993$, $-2.001$, $-1.995$
for $(\sigma,\mu)=(0.1,0)$, $(0.1,-0.05)$, $(0.05,0)$, respectively.
These results suggest that the scalar field decays as $v^{-2}$ along
the future horizon. 
To determine the coefficient of the power law decay,
we fit the $\phi|_{r=M}$ to the function $\alpha H_1/v^2$ for $1000\leq v \leq 2000$.
Then we find $\alpha=0.659$, $0.665$, $0.661$ 
for $(\sigma,\mu)=(0.1,0)$, $(0.1,-0.05)$, $(0.05,0)$, respectively.
Hence, for all cases that we have considered, the late time behaviour of $\phi|_{r=M}$ is given by\begin{equation}
\label{latel1}
 \phi|_{r=M} \sim \frac{0.66 H_1}{v^2}  \qquad \qquad v\to \infty\ .
\end{equation}
We will see in section \ref{AdS2} that Eq.~(\ref{latel1}) can be obtained from an $AdS_2$ calculation. In Ref. \cite{Aretakis:2011hc}
it was shown analytically that the $l=1$ mode of the scalar field decays at least as fast as $v^{-3/4}$ for large $v$; this is consistent with our numerical result.
As we will explain below, we expect the coefficient of $v^{-2}$ to be modified for initial data with non-vanishing NP constant.

\subsubsection{Zero Aretakis constant}

\begin{figure}
  \centering
  \subfigure
  {\includegraphics[scale=0.45]{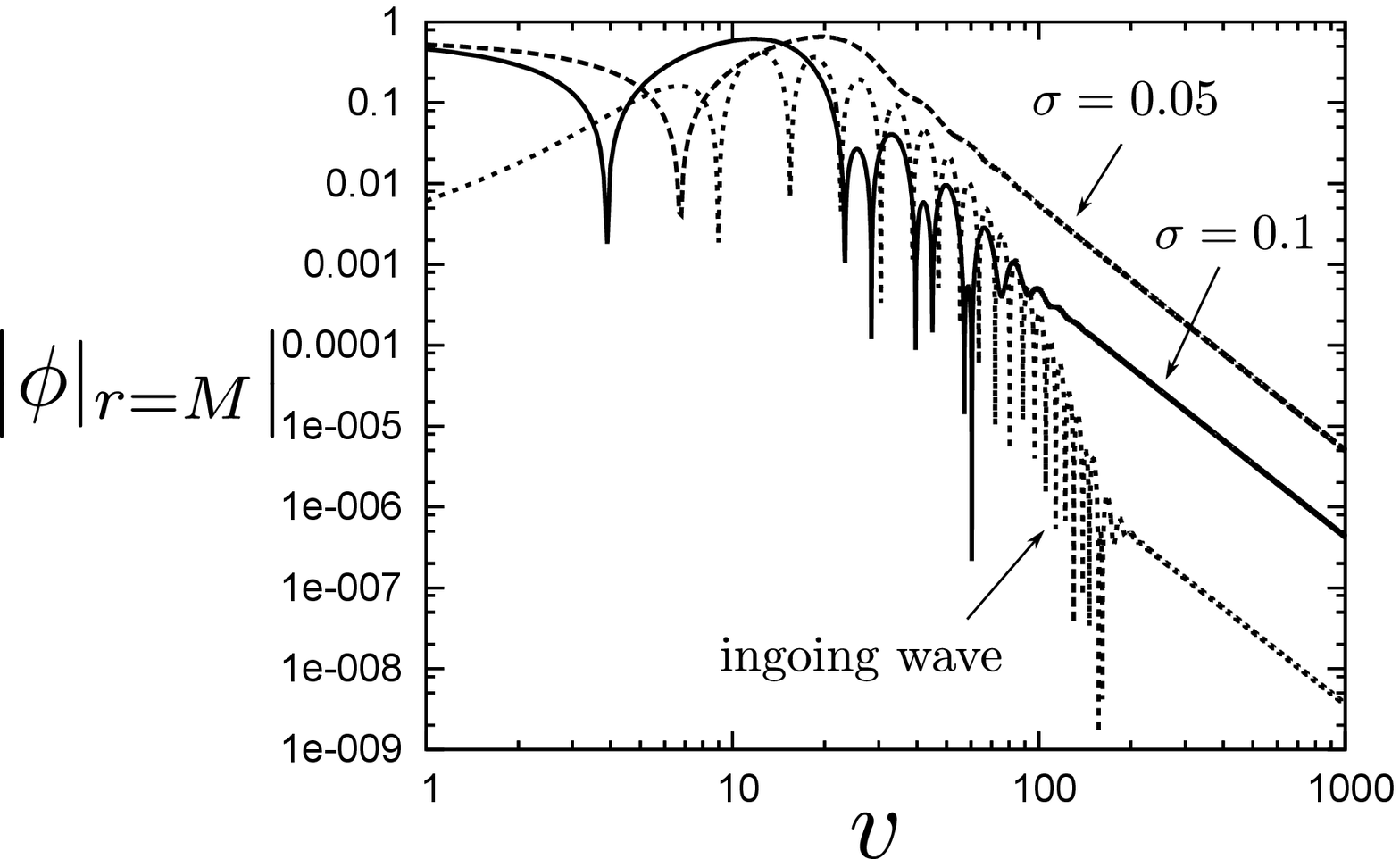}
\label{phiH_l1I0}
  }
  \subfigure
  {\includegraphics[scale=0.45]{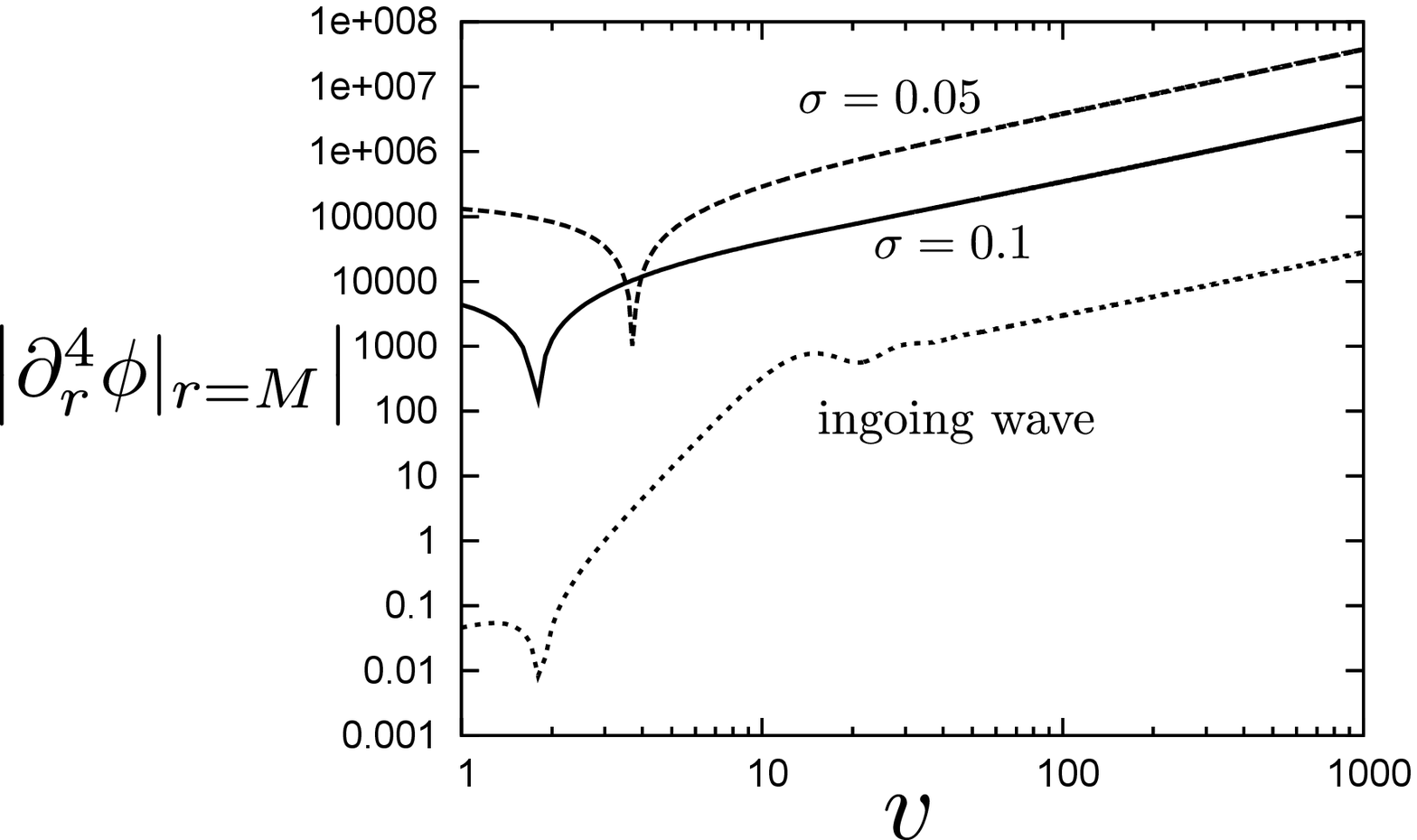} 
\label{phirrrrH_l1I0}
  }
  \caption{
Time evolution of $\phi|_{r=M}$ and $\partial_r^4\phi|_{r=M}$ for
 various initial data with $l=1$ and $H_1=0$: outgoing wavepackets with
$\sigma=0.05,0.1$, $\mu=\sigma(\sigma-\sqrt{\sigma^2+4M^2})/(2M)$ and an ingoing wavepacket with $(\sigma',\mu')=(3.0,10.0)$. 
We find that $\phi|_{r=M}$ decays and $\partial_r^4\phi|_{r=M}$ blows up linearly in $v$. 
\label{phiH_l1I0_all}
}
\end{figure}

Now consider the case $l=1$ with $H_1=0$.
For the outgoing wavepacket
we choose the parameters $\sigma=0.05, 0.1$ and 
$\mu=\sigma(\sigma-\sqrt{\sigma^2+4M^2})/(2M)$ to set $H_1=0$.
For the ingoing wave we choose the parameters 
$(\sigma',\mu')=(3.0,10.0)$.
We plot the time evolution of $\phi|_{r=M}$ and
$\partial_r^4\phi|_{r=M}$ in
Fig.~\ref{phiH_l1I0_all}.
This is  consistent with the following behaviour: $\phi|_{r=M}$ decays and $\partial_r^4\phi|_{r=M}$ blows up linearly in $v$. Thus, we can conclude that there is an instability for $l=1$ even when the Aretakis constant vanishes. However, just as for $l=0$, it affects quantities with one more $r$-derivative than in the case with non-vanishing Aretakis constant.

Now consider the late time behaviour of $\phi|_{r=M}$.
Fitting the absolute value of $\phi|_{r=M}$ to a power law decay $v^{a}$, 
we obtain the exponents $a=-3.001$, $-2.990$ for the
outgoing wave with $\sigma=0.05, 0.1$ and $a=-2.990$ for the ingoing wave, respectively.
This suggests  that, for $H_1=0$, the late
time behaviour of $\phi|_{r=M}$ is given by
\begin{equation}
 \phi|_{r=M}\sim \frac{C}{v^3} \qquad
v\to\infty\ ,
\end{equation}
for some constant $C$ depending on the initial data.

\subsection{Late time tails outside the horizon}

\begin{figure}
\begin{center}
\includegraphics[scale=0.5]{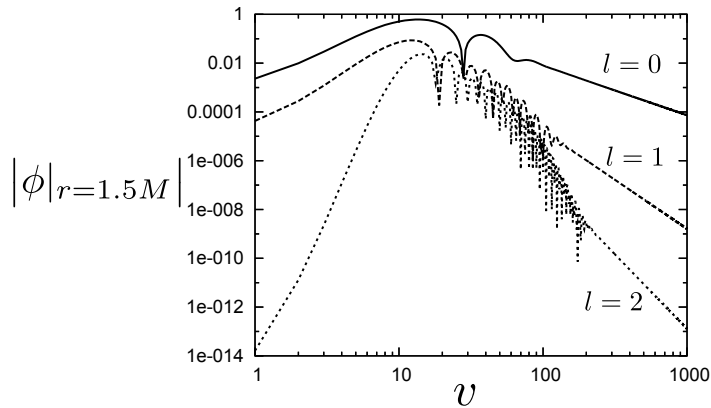}
\end{center}
\caption{
Time dependence of scalar field $\phi$ outside the horizon: $r=1.5M$.
We consider outgoing wavepackets with 
$(\sigma,\mu)=(0.1,-0.1),(0.1,0),(0.05,-0.05)$
for $l=0,1,2$, respectively. For such initial data, the conserved
 quantities are non-zero. After an initial period of damped oscillation (quasinormal ringing) we find power law decay $\phi|_{r=1.5M}\sim C v^{-2l-2}$.
}
 \label{phiout}
\end{figure}

Now we consider the late time behaviour of the scalar field {\it outside} the horizon. We consider outgoing wavepackets with $(\sigma,\mu)=(0.1,-0.1),(0.1,0),(0.05,-0.05)$
for $l=0,1,2$, respectively. In Fig.~\ref{phiout} we plot the time dependence of $\phi$ at
$r=1.5M$. Initially the field is supported near the horizon ($r=M$) so the amplitude at $r=1.5M$ is small. The amplitude grows as the field disperses outwards from the black hole, and then decays as the field disperses to infinity. 

Our results are consistent with power law decay at late time.
Fitting to $v^{a}$ we obtain the exponents  $a=-2.008$, $-4.006$, $-6.026$ for $l=0,1,2$, respectively.
For the fitting, we use numerical data in the range $1000\leq v \leq 2000$
for $l=0,1$. For $l=2$, we used data in the range $800\leq v \leq 1000$
since the numerical calculation breaks down at $v\geq 1000$. (We observed
small oscillations which depend on the resolution.)

From these result, we extrapolate that the late time behaviour of
the $l$th partial wave of the scalar field outside the horizon is given by
\begin{equation}
\label{tail}
 \phi|_{r=r_0} \sim C v^{-2l-2} \qquad \qquad v \to \infty
\end{equation}
where $r_0>M$ for some constant $C$ depending on $r_0$ and the initial data. We will reproduce this power-law from an $AdS_2$ calculation in the next section.\footnote{
The decay at fixed $r$ outside the horizon involves a different power of $v$ than the decay on the horizon. This is consistent: $C$ will diverge as $r \rightarrow M$.}
 
Ref. \cite{Blaksley:2007ak} argued that the decay outside the horizon should be proportional to $v^{-(2l+\mu+1)}$ where $\mu=1$ if the NP constant is non-zero and $\mu=2$ otherwise.\footnote{
Ref. \cite{Blaksley:2007ak} said that $\mu=1$ for an ``initially static moment'' and $\mu=2$ otherwise. But, according to Ref. \cite{Gomez:1994ns}, the precise notion of an ``initially static moment" is a non-vanishing NP constant.} Our results show that this is not quite correct: instead we should have $\mu=1$ if {\it either} the Aretakis constant {\it or} the NP constant is non-zero (Ref.  \cite{Blaksley:2007ak} considered only initial data with vanishing Aretakis constant.) This makes sense given the existence of the conformal isometry discussed in section \ref{CI}, which interchanges a solution with non-vanishing Aretakis constant with one with non-vanishing NP constant. 


Ref. \cite{Blaksley:2007ak} found that the decay along the horizon is proportional to $v^{-(l+1)}$ for vanishing Aretakis constant but non-vanishing NP constant. This implies that, in general, the coefficient of $v^{-1}$ in (\ref{latephi}) and $v^{-2}$ in (\ref{latel1}) will contain an extra term proportional to the NP constant (which is zero for our initial data). The decay along the horizon is proportional to $v^{-(l+\mu)}$, as found in Ref. \cite{Blaksley:2007ak}, but with $\mu$ defined as above.

\section{Scalar fields in $AdS_2$}
\label{AdS2}

In this section we will show that certain features of the late time behaviour on and outside the horizon of extreme RN can be deduced by looking purely at the near-horizon geometry.

\subsection{Coordinates}

\begin{figure}
\begin{center}
\includegraphics[scale=0.4]{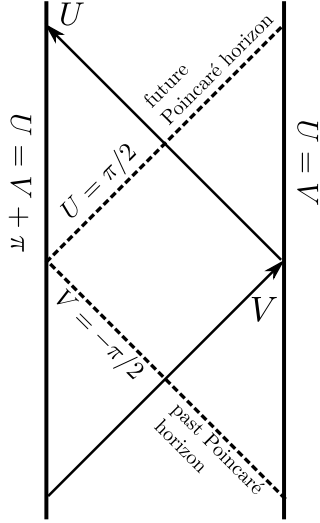}
\end{center}
\caption{
Penrose diagram for $AdS_2$.
}
 \label{penroseAdS2}
\end{figure}

The near-horizon limit of extreme RN is $AdS_2 \times S^2$ where each factor has radius $M$. Taking $M=1$, the $AdS_2$ metric in static coordinates is
\be
 ds^2 = -r^2 dt^2 + \frac{dr^2}{r^2}  \; .
\ee 
Defining
\be
 u = t+\frac{1}{r}, \qquad v = t - \frac{1}{r}
\ee
we obtain the metric in ingoing Eddington-Finkelstein coordinates 
\be
 ds^2 = -r^2 dv^2 + 2 dv dr  \; .
\ee
These coordinates are regular at the future Poincar\'e horizon $r=0$. We can also use $(u,v)$ as coordinates:
\be
 ds^2 = -\frac{4}{(u-v)^2} du dv  \; .
\ee
Now define Kruskal-like coordinates
\be
U = \tan^{-1} u, \qquad V = \tan^{-1} v
\ee
to obtain 
\be
 ds^2 = -\frac{4}{\sin^2 (U-V)} dU dV  \; .
\ee
These are global coordinates on $AdS_2$. The Penrose diagram for $AdS_2$
is depicted in Fig.~\ref{penroseAdS2}.
The region covered by the
static coordinates is $-\pi/2< V < U < \pi/2$. The left timelike
infinity is $U=V+\pi$, the right timelike infinity is $U=V$. The future
Poincar\'e horizon of the static patch is $U=\pi/2$, the past Poincar\'e
horizon is $V = - \pi/2$. 

\subsection{Massless scalar in $AdS_2$}

Consider first a massless scalar $\psi$ field in $AdS_2$. Since its equation of motion is conformally invariant, we can write down the general solution: 
\be
\psi = f(U) + g(V)
\ee
for arbitrary functions $f,g$. To compare with our earlier results, we need to calculate the Aretakis conserved quantity. Writing out the wave equation in $(v,r)$ coordinates and evaluating at $r=0$ (or taking the near-horizon limit of (\ref{H0def})) shows that the following quantity is conserved along the future Poincar\'e horizon
\be
 H_0[\psi] \equiv (\partial_r \psi)_{r=0} \; .
\ee
Let's evaluate this in $(U,V)$ coordinates. Converting from $(v,r)$ coordinates to $(U,V)$ coordinates gives
\be
 \frac{\partial}{\partial r} = - \frac{\sin^2(U-V)}{2\cos^2 V} \frac{\partial}{\partial U} \; .
\ee
Hence
\be
 H_0  =- \frac{1}{2} f'(\pi/2)  \; .
\ee
Now consider the second $r$-derivative of the field at the horizon:
\be
 (\partial_r^2 \psi)_{r=0} = \frac{1}{2}  f'(\pi/2)\tan V +\frac{1}{4} f''(\pi/2) = -H_0 v +\frac{1}{4} f''(\pi/2)
\ee
so when $H_0 \ne 0$ we have linear growth in $v$, just as for extreme RN. On the other hand if $H_0=0$ then we see that $\partial_r^2 \psi |_{r=0}$ is a constant; by taking another $r$-derivative we see that if this constant is non-zero then $\partial_r^3 \psi |_{r=0}$ will blow up linearly in $v$, just as we found numerically for extreme RN.

Is this an instability of the scalar field in $AdS_2$? No: working in $(U,V)$ coordinates one sees that $\psi$ and all of its derivatives are bounded. The blow-up in $(v,r)$ coordinates is simply a coordinate effect, it has no invariant significance. By contrast, in extreme RN, the $(v,r)$ coordinates are a preferred set of coordinates that follow from the symmetries of the spacetime. The function $r$ is defined using spherical symmetry and there is a unique Killing vector field $K^\mu$ that is timelike and canonically normalized ($K^2 = -1$) at infinity which is used to define $v$. In $AdS_2$ there exist infinitely many timelike Killing vector fields analogous to $K^\mu$. Hence, unless one has some reason to regard one of these as preferred, then there is no invariant way of introducing $(v,r)$ coordinates and hence no instability.

Now consider the late time behaviour of the field. The value of the field on the Poincar\'e horizon is
\be
 \psi(\pi/2,V) = f(\pi/2)+g(V)  \; .
\ee
Converting to $(v,r)$ coordinates, the late time behaviour of the field at the horizon ($v \rightarrow \infty$, $r=0$) is
\be
 \psi(\pi/2,V) = f(\pi/2) + g(\pi/2) - \frac{ g'(\pi/2) }{v} +  {\cal O}\left( \frac{1}{v^2} \right)  \; .
\ee
So far we have imposed no boundary condition on $\psi$. Note that the value of $\psi$ on the right timelike infinity is\footnote{
In the language of AdS/CFT, $\psi_\infty(U)$ is the source for the operator dual to $\psi$.}
 \be
 \psi_\infty(U) \equiv f(U) + g(U)
\ee
where we use $U$ as the coordinate at right timelike infinity. We now impose the boundary condition that $\psi_\infty(U)$ and $\psi_\infty'(U)$ decay at late time at right timelike infinity:
\be
\label{latetimebc}
 \psi_\infty(U) \to 0, \qquad  \psi_\infty'(U) \rightarrow 0, \qquad {\rm as} \qquad U \rightarrow \pi/2 \; ,
\ee
which is equivalent to
\be
 g(\pi/2) = - f(\pi/2), \qquad g'(\pi/2)= -f'(\pi/2)
\ee
and hence
\be
 \psi(\pi/2,V) = -\frac{2H_0}{v} + {\cal O}\left( \frac{1}{v^2} \right) \; .
\ee
This agrees to good accuracy with our numerical results for extreme RN (summarized in equation (\ref{latephi})).\footnote{Recall that this result applies only for the case of vanishing NP constant. If the NP constant is non-zero then there is radiation present near infinity in extreme RN. Maybe one could reproduce results for this case by modifying the boundary condition for $\psi_\infty'$ in (\ref{latetimebc}). Note also that we do not obtain the $\ln v/v^2$ corrections of (\ref{latephi_second}) from this $AdS_2$ calculation.} The late time behaviour outside the horizon ($v \rightarrow \infty$ at fixed $r>0$) is, if we make the extra assumption $\psi_\infty''(U) \rightarrow 0$ as $U \rightarrow \pi/2$, 
\be
 \psi =   -\frac{4 H_0}{rv^2} + {\cal O}\left( \frac{1}{v^3} \right) \; .
\ee
This is in good agreement with (\ref{tail}). Note that our assumptions on $\psi_\infty$ are all satisfied if there exists $U_0<\pi/2$ such that $\psi_\infty$ vanishes for $U>U_0$.

\subsection{Massive scalar in $AdS_2$}
\label{Sec:massiveAdS2}

Now consider a scalar field $\psi$ of mass $m$ in $AdS_2$. Since the general solution of the wave equation is not so simple in the massive case, we will proceed more heuristically than we did above. We assume that, in a neighbourhood of $U=V=\pi/2$ (i.e.\ late time at right timelike infinity) the field satisfies ``normalizable" boundary conditions
\be
\label{massiveads2bc}
\psi = (U-V)^{\Delta_m}F(V) + {\cal O}((U-V)^{\Delta_{m}+1})
\ee
where $F(V)$ is an arbitrary function and
\be
\label{conformalweight}
 \Delta_m \equiv \frac{1}{2} + \sqrt{m^2 + \frac{1}{4}} \; .
\ee
Taking the near-horizon limit of the $l$th multipole $\psi_l$ of a {\it massless} scalar in extreme RN (setting $M=1$) results in a scalar field of mass $m$ in $AdS_2$ where
\be
m^2 = l(l+1), \qquad l=0,1,2,\ldots
\ee
and therefore $\Delta_m= l+1$.
In this case, the conserved quantity at the Poincar\'e horizon is simply
\be
 H_l[\psi] = (\partial_r^{l+1} \psi)_{r=0} \; .
\ee
We can now calculate $H_l[\psi]$ at large $v$: $r=0$, $v\rightarrow \infty$ is equivalent to $U=\pi/2$, $V \rightarrow \pi/2$, and we can use the behaviour (\ref{massiveads2bc}) of $\psi$ near the boundary to obtain
\be
 H_l = \frac{(-1)^{l+1} (2l+1)!}{2^{l+1} l!} F(\pi/2) \; .
\ee
Hence, near $U=V=\pi/2$ we have
\be
 \psi = \frac{l!}{(2l+1)!} H_l \left( -2(U-V) \right)^{l+1} + {\cal O}((U-V)^{l+2}) \; .
\ee
From this expression we can calculate the late time behaviour along the Poincar\'e horizon in $(v,r)$ coordinates ($r=0$, $v \rightarrow \infty$):
\be
 \psi = \frac{l!}{(2l+1)!} H_l \left( -\frac{2}{v} \right)^{l+1} + {\cal O}\left( \frac{1}{v^{l+2}} \right) \; .
\ee 
For $l=1$ this agrees to good accuracy with our numerical result (summarised in (\ref{latel1})) for the full extreme RN geometry. The late time behaviour outside the horizon ($r>0$ fixed, $v \rightarrow \infty$):\footnote{
It follows that, in AdS/CFT, the conserved quantity $H_l$ determines the late time behaviour of the expectation value of the operator dual to $\psi$.
}
\be
 \psi = \frac{l!}{(2l+1)!}H_l \left( -\frac{4}{rv^2} \right)^{l+1} + {\cal O}\left( \frac{1}{v^{2l+3}} \right)
\ee
in agreement with (\ref{tail}). 

Why are these $AdS_2$ results in such good agreement with our numerical results for the full extreme RN geometry? One might have thought that the late time behaviour at the horizon would exhibit some dependence on the geometry outside the horizon, which is different in $AdS_2$ and extreme RN. Maybe the numerical coefficients in  (\ref{latephi}) and (\ref{latel1}) would deviate from the $AdS_2$ values if they were computed to higher accuracy. Or maybe all that matters at late time is the geometry exactly at the horizon, which would explain the agreement. It would be nice to understand this better.

Observe that taking the near-horizon limit of a spherically symmetric scalar of mass $m$ in extreme RN (again with $M=1$) also results in a scalar of mass $m$ in $AdS_2$.
From the above we deduce that if $m^2=n(n+1)$ where $n$ is a positive integer the above calculations remain valid (with the replacement $l \to n$). In particular, we learn there are conserved quantities  for certain {\it massive} scalar fields in the near-horizon geometry of extreme RN. In Sec.~\ref{massiveconstants} we show that in fact a massive scalar field in the full extreme RN also possesses such conserved quantities, for these specific values of the mass, and show they lead to instabilities.

For general $m$ it does not appear possible to construct conserved quantities on the horizon; however from (\ref{massiveads2bc}) we may still deduce the late time behaviour.  As $v \to \infty$  we find that on the horizon
\be
(\partial_r^{k} \psi)_{r=0} \sim \frac{(-1)^{k} \Gamma(\Delta_m+k) F(\pi/2)}{2^{k}\Gamma(\Delta_m)} \, v^{k-\Delta_m}  
\ee
for all $k \geq 0$: we see this blows up for $k \geq \lfloor \Delta_m  \rfloor +1$, where $\lfloor~\rfloor$ denotes the integer part. In Sec.~\ref{Nummassisve} we will demonstrate numerically that instabilities of this nature indeed occur in the full extreme RN for certain $m$ such that $\Delta_m$ is not integer (despite the absence of a conserved quantity). For completeness, the late time behaviour outside the horizon at fixed $r>0$ is
\be
\psi \sim  \frac{F(\pi/2)}{(rv^2)^{\Delta_m}}  \; .
\ee


\section{Massive scalar field}
\label{Nummassisve}

In this section we consider the evolution of a spherically symmetric massive scalar field in the extreme RN background. First we will show that, for special values of the mass of the scalar field, a tower of conserved quantities exists just as for the massless scalar field. These can be used to deduce the existence of instabilities. Then we will solve the equation of motion numerically as in the massless case to investigate the evolution in more detail.

\subsection{Conserved quantities on the horizon}
\label{massiveconstants}

We start by writing down the massive wave equation (\ref{KG}) for a spherically symmetric massive scalar field in ingoing EF coordinates
\begin{equation}
 2 r \partial_v \partial_r (r\psi) 
+ \partial_r \left( \Delta \partial_r \psi \right) 
- m^2 r^2 \psi= 0 \,.
\end{equation}
We find that the construction of the conserved quantities for the massless scalar field~(\ref{H_l}) can be generalised to a massive scalar for special values of the mass: 
\be
m^2 = n(n+1)M^{-2}
\ee
where $n$ is a positive integer. This gives a very similar term in the equation of motion to the centrifugal barrier of a massless scalar field, so the construction of conserved quantities proceeds analogously to that of the massless scalar with $n \leftrightarrow l$. Explicitly, we find 
\begin{equation}
H_n[\psi] \equiv 
\frac1M
\left\{
\partial_r^n \left[
f_n(r) \partial_r (r\psi)
\right]\right\}_{r=M}
\label{Hl_massive}
\end{equation}
is conserved along $\mathcal{H}^+$,
where $f_n(r)$ is a polynomial defined by\footnote{In fact any smooth $f_n(r)$ whose derivatives $f_n^{(k)}|_{r=M}$, for $k=0, \dots, n$, coincide with those of this polynomial will do.}
\begin{equation}
f_n(r) 
\equiv
\sum_{k=0}^n c_k \left(\frac{r}{M}-1\right)^k
\end{equation}
whose coefficients are recursively determined by $c_0=1, c_1=-n$ and for $2 \leq k\leq  n$
\begin{equation}
c_{k} = -\frac{n(n+1)
\left[
2 ( n - k + 1 ) c_{k-1} + ( n - k ) c_{k-2}
\right]
}{ k (n-k+2)( 2n - k + 1 ) }
\, .
\end{equation}
It follows that generically not all of the quantities $\partial_r^{j}\psi$ for $0 \le j \le n+1$ can decay at late time on ${\cal H}^+$. By analogy with the $l>0$ massless scalar, we expect, and our numerics confirms, that $\partial_r^{j}\psi$ decays at late time on ${\cal H}^+$ for $j\leq n$. Then $\partial_r^{n+1}\psi$ does not decay and $\partial_r^{n+2}\psi$ blows up linearly along ${\cal H}^+$.

It is interesting to ask what happens for general mass $m$. Although we do not find any conservation laws in this case, by continuity in the mass parameter one might expect that, generically, $\partial_r^{n+1} \psi$  does not decay, and $\partial_r^{n+2} \psi$ blows up, at late time on ${\cal H}^+$ where $n$ is given by
\begin{equation}
n = \lfloor \Delta_m \rfloor  - 1
\label{ngen}
\end{equation}
where $\Delta_m$ is given by Eq.~(\ref{conformalweight}) with $m$ replaced by $Mm$ and $\lfloor~\rfloor$ represents the integer part. We confirm this expectation numerically in the next section.

\subsection{Numerical results}

Consider the massive scalar wave equation in the coordinates developed earlier~(\ref{waveUv}). Using the numerical method introduced for the massless scalar field discussed in section \ref{Nummassless}, 
we study the time dependence of a
massive scalar field.

\subsubsection{Non-zero conserved quantity}

In this section, we study the time evolution for the case corresponding to that studied in 
section \ref{Sec:massless_nonzero}, 
for which the conserved quantities~(\ref{Hl_massive}) are non-zero.

\begin{figure}[htbp]
\centering
\subfigure[$\phi(v,r)$]{
\includegraphics[width=7.5cm, clip]{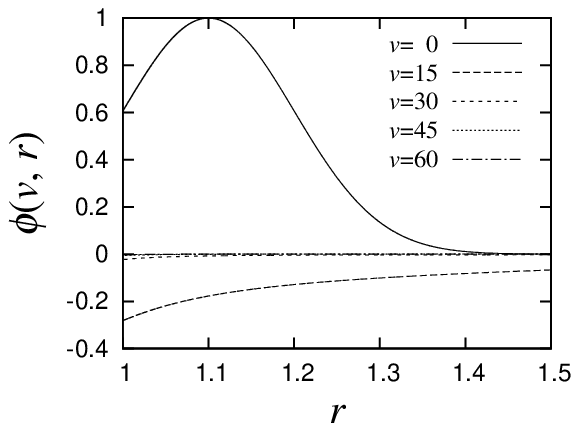}
}
\subfigure[$H_1(v,r)$]{
\includegraphics[width=7.5cm, clip]{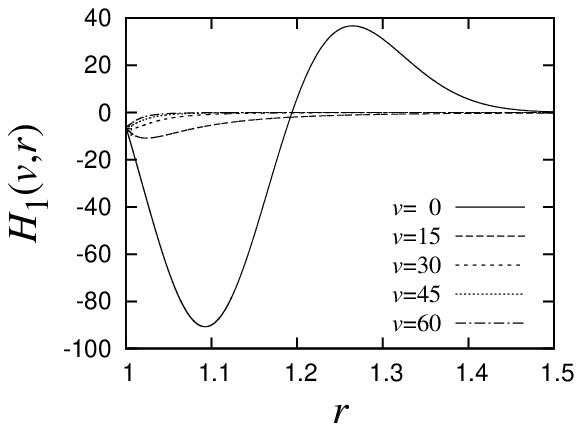}
}
\caption{$\phi(v,r)$ and $H_1(v,r)$ for the massive scalar field with
$(Mm)^2 = 2$.
}
\label{r-phi_m2}
\end{figure}

First of all, we show the time evolution of $\phi(v,r)=r\psi(v,r)$ on and outside the horizon
for $n=1$ ($(Mm)^2 = 2 $)
using the outgoing wave initial data~(\ref{out}) 
with $(\sigma,\mu)=(0.1, -0.1)$.
In Fig.~\ref{r-phi_m2}, we plot $\phi(v,r)$ and the quantity $H_1(v,r)$ defined by
\begin{equation}
H_1(v,r) \equiv \frac1M \partial_r\left[
f_1(r) \partial_r \phi
\right]
= \frac1M \left[
\left(2-\frac{r}{M}\right)\partial_r^2 \phi 
- \frac1M \partial_r \phi
\right],
\end{equation}
which coincides with the conserved quantity~(\ref{Hl_massive}) for 
$n=1$ at the horizon. 
Panel~(a) shows that $\phi(v,r)$ quickly decays on and outside the horizon. 
Panel~(b) shows that $H_1(v,r)$ stays constant on the horizon while it decays 
to zero outside the horizon. 
This property is consistent with $\partial_r^2\phi|_{r=M}$ being constant and $\partial_r^3\phi|_{r=M}$ diverging linearly in $v$ at late time.

\begin{figure}[htbp]
\centering
\subfigure[$\left|\phi|_{r=M}\right|$]{
\includegraphics[width=7.5cm, clip]{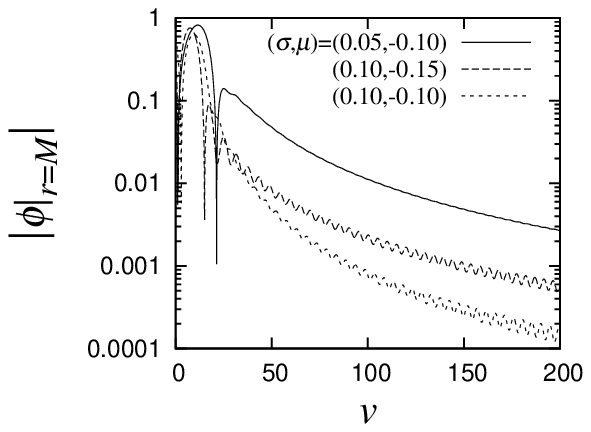}
}
\subfigure[$\left|\partial_r^3 \phi|_{r=M}\right|$]{
\includegraphics[width=7.5cm, clip]{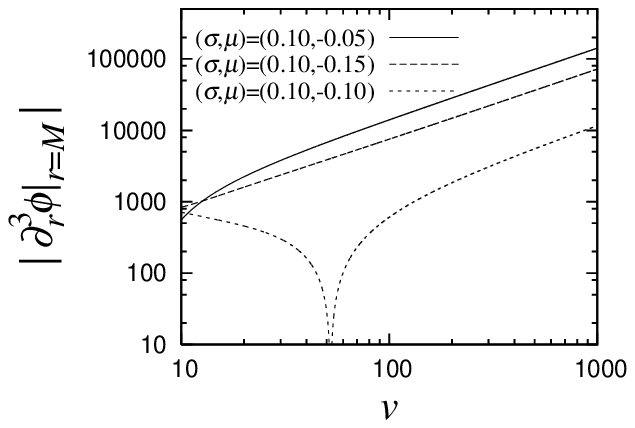}
}
\caption{
Time evolution of $\phi|_{r=M}$ and $\partial^3_r \phi|_{r=M}$ at the horizon
for $(Mm)^2 = 2$.
}
\label{phiH_m2}
\end{figure}

In Fig.~\ref{phiH_m2}, we show $\phi|_{r=M}$ and $\partial_r^3 \phi|_{r=M}$ for 
$(Mm)^2 = 2$ for various initial data with $H_1\neq 0$.
To generate this plot, we used the outgoing initial data~(\ref{out}). Panel~(b) shows that $\partial^3_r\phi|_{r=M}$ is linearly divergent with respect to time,
as expected from existence of the conserved quantity $H_1$ (this could be proved analytically in the same way as for the massless scalar with $l=1$). 

In Panel~(a),
we find that, at late time, $\phi|_{r=M}(v)$ shows oscillation with frequency $\omega = m$.  Oscillations with this frequency have been seen in previous studies of massive scalar fields in black hole backgrounds~\cite{Koyama:2000hj}. It is convenient to separate the value of the field at the horizon into its mean value and an oscillating component:
\begin{equation}
\phi|_{r=M}(v) =\left. \phi^\text{mean}(v) + \phi^\text{osci}(v) \right|_{r=M} \,,
\end{equation}
where $\phi|_{r=M}^\text{mean}(v)\equiv T^{-1} \int^{v+T/2}_{v-T/2}\phi|_{r=M}(v') dv'$ with $T=2\pi/m$. 
At late time, the oscillating component decays more slowly than the mean value. 
On the other hand, we observed that for $\partial_r^{k\geq 1}\phi|_{r=M}$, the oscillations are subdominant at late time for generic initial data. This suggests that the instability is associated to $\phi^\text{mean}$.

\begin{figure}[htbp]
\centering
\includegraphics[width=7.5cm, clip]{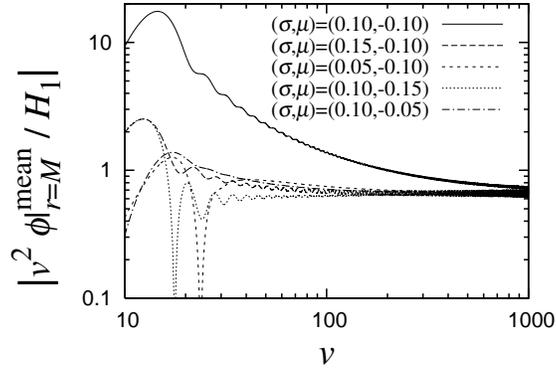}
\caption{$\left|v^2 \phi|_{r=M}^\text{mean} / H_1\right|$ for outgoing wave initial data for 
$(\sigma, \mu) = (0.1, -0.1)$, 
$(0.15, -0.1)$,
$(0.05, -0.1)$,
$(0.1, -0.15)$ and
$(0.1, -0.05)$.
For any initial data, 
$|v^2 \phi|_{r=M}^\text{mean} / H_1|$ tends to $0.66$ at late time.
}
\label{v2phi_m2}
\end{figure}

Next we examine in more detail the behaviour of each component in $\phi|_{r=M}$.
First, we show the behaviour of $\phi|_{r=M}^\text{mean}(v)$ for $(Mm)^2 = 2$
in Fig.~\ref{v2phi_m2}.
We find that $\phi|_{r=M}^\text{mean}(v)$ decays by a power law, and its value is 
governed by the conserved quantity $H_1$ as
\begin{equation}
\phi|_{r=M}^\text{mean} \sim \frac{0.66H_1}{v^2}\,,
\label{eq:v2phi_m2}
\end{equation}
for several different choices of outgoing wave initial data. 
More precisely, we calculated the late-time behaviour of $\phi|_{r=M}$ for outgoing wave 
initial data for $(\sigma, \mu) = (0.1, -0.1)$, 
$(0.15, -0.1)$,
$(0.05, -0.1)$,
$(0.1, -0.15)$ and
$(0.1, -0.05)$.
%
We fitted $\phi|_{r=M}^\text{mean} $ to the function $\alpha H_1/v^2$ for $1000\leq v \leq 2000$,
and found $\alpha=0.715$, $0.662$, $0.668$, $0.658$ and $0.665$ respectively for these initial data.
We also found that $\alpha$ for $(\sigma, \mu) = (0.1, -0.1)$ becomes $0.657$
if we change the fitting function into $\alpha H_1/v^2 + \beta \log v / v^3$ 
introducing an ansatz for the subleading term.
Note that the behaviour of Eq.~(\ref{eq:v2phi_m2})
was seen also in the case of massless scalar field with $l=1$
(See Eq.~(\ref{latel1})).

\begin{figure}[htbp]
\centering
\subfigure[$A|^\text{osci}_{r=M}(v)$]{
\includegraphics[width=7.5cm, clip]{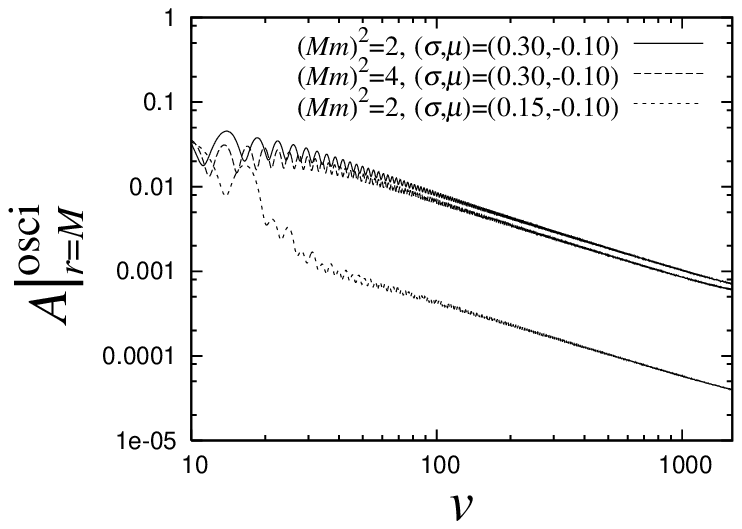}
}
\subfigure[$A|^\text{osci}_{r=1.5M}(v)$]{
\includegraphics[width=7.5cm, clip]{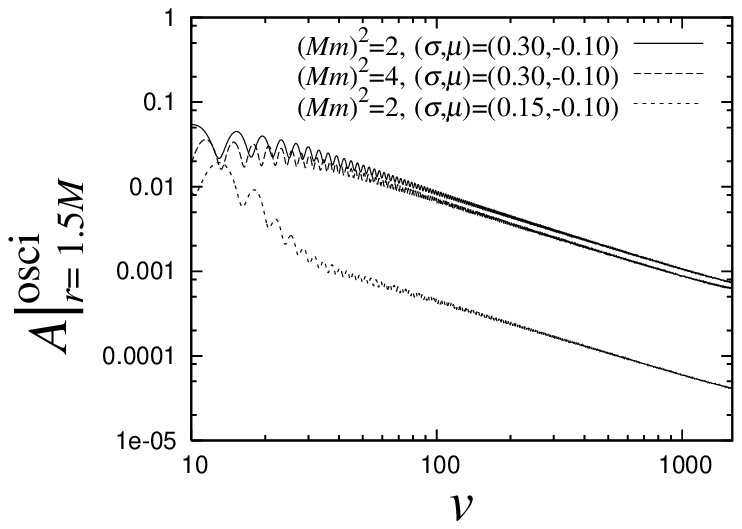}
}
\caption{
Panel~(a):
Amplitude of $\phi|_{r=M}^\text{osci}$ and 
$A|^\text{osci}_{r=M}(v)$,
in the case 
of outgoing wave initial data with
$(\sigma, \mu) = (0.3, -0.1)$
for $(Mm)^2=2, 4$ and that with
$(0.15, -0.11)$ for $(Mm)^2=2$.
Panel~(b): $A|^\text{osci}(v)$ at $r=1.5M$ for the same initial data.
Numerical results suggest that
$A|^\text{osci}_{r=M}(v)$ and $A|^\text{osci}_{r=1.5M}(v)$ 
become proportional to $v^{-5/6}$ at late time for any initial data.}
\label{norm_m2}
\end{figure}

Next, we examine the behaviour of $\phi|_{r=M}^\text{osci}$.
Since this component oscillates with frequency $\omega=m$ at late time, 
it is useful to focus on its amplitude 
$A|_{r=M}^\text{osci}(v)$ defined by
\begin{equation}
A|_{r=M}^\text{osci} (v) 
\equiv 
\sqrt{
\phi^2 + m^{-2} (\partial_v\phi)^2
}\Big|_{r=M}^\text{osci}\,.
\end{equation}
%
In Panel~(a) of Fig.~\ref{norm_m2}, we plot 
$A|^\text{osci}_{r=M}(v)$ in the case of outgoing wave initial data with
$(\sigma, \mu) = (0.3, -0.1)$ for $(Mm)^2=2, 4$ and with $(0.15, -0.1)$ for $(Mm)^2=2$.
Fitting the data to the function $Cv^a$, we find the power of decay is
$a=-0.84$, $0.83$ and $0.82$ for our different initial data.
These results are consistent with the scaling $A|_{r=M}^\text{osci} (v) \propto v^{-5/6}$.
We plot also the oscillation amplitude of $\phi(v,r)$ {\it outside} the horizon in Panel~(b)
for the same initial data,
which shows the same scaling $A|_{r=1.5M}^\text{osci} (v) \propto v^{-5/6}$.
More precisely, fitting the late time behaviour with $Cv^a$ gives $a=-0.84$, $0.84$ and $0.82$
for our different initial data.
Note that $v^{-5/6}$ scaling for the late-time tail of a massive scalar field outside a black hole horizon has been discussed previously in Refs.~\cite{Koyama:2000hj,Burko:2004jn}. The constant $C$ in these fits does not appear to be simply related to the conserved quantity $H_1$.

\begin{figure}[htbp]
\centering
\subfigure[$H_0$]{
\includegraphics[width=7.5cm, clip]{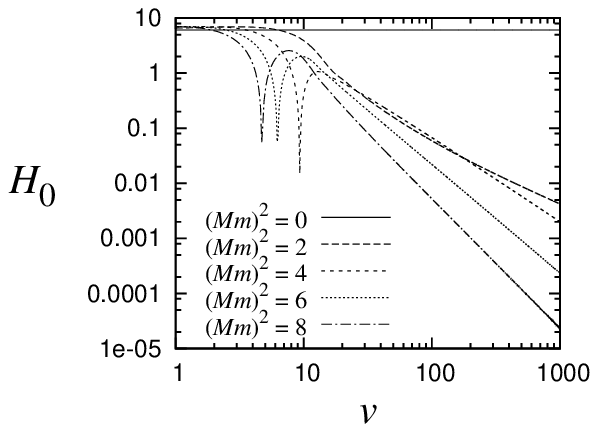}
}
\subfigure[$H_1$]{
\includegraphics[width=7.5cm, clip]{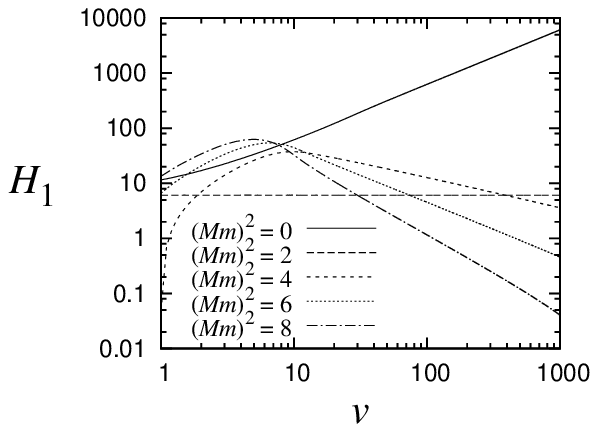}
}
\caption{
$m^2$ dependence of
$H_0(v)$ and $H_1(v)$ for $(Mm)^2=0, 2, \ldots, 8$.}
\label{mvar_H}
\end{figure}

\begin{figure}[htbp]
\centering
\subfigure[$(Mm)^2$ v.s.\ $p_0$]{
\includegraphics[width=7.5cm, clip]{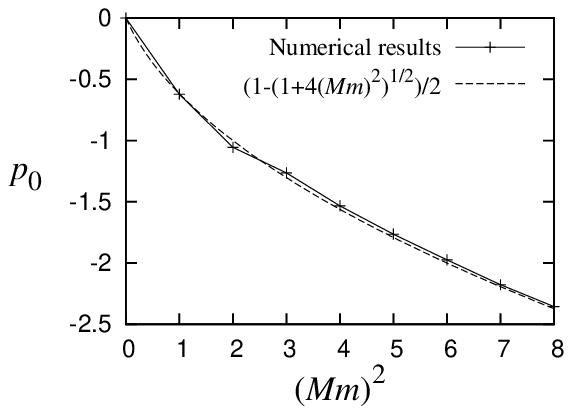}
}
\subfigure[$(Mm)^2$ v.s.\ $p_1$]{
\includegraphics[width=7.5cm, clip]{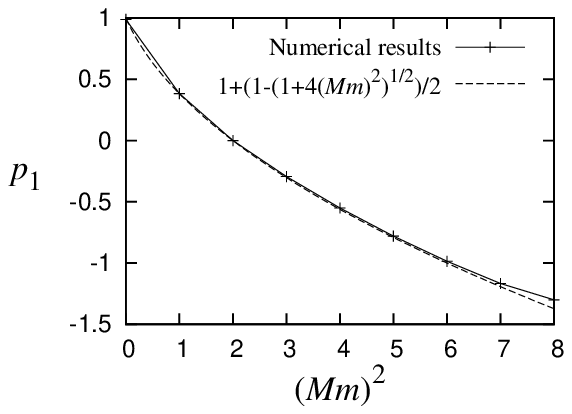}
}
\caption{
$m^2$ dependence of $p_n$, where we assume 
$H_n \propto v^{p_n}$.
In both panels, the
numerically obtained $p_n$ for 
$(Mm)^2=0, 1, \ldots, 8$ are shown
with analytic curves 
$p_n = n + \bigl(1-\sqrt{1+4(Mm)^2}\bigr)/2$.
For any data point, relative error of the numerical value
from the analytic value is $\sim 5\times 10^{-2}$ at most.
}
\label{pOFm2}
\end{figure}

We also comment on the dependence of $\phi|_{r=M}$ on the scalar field mass $m$.
For this purpose, it is useful to focus on time dependence of 
$H_0$ and $H_1$, which are conserved when $(Mm)^2 = 0$ and $2$, respectively.
We summarize their time dependence in Fig.~\ref{mvar_H}.
These quantities are dominated by $\phi^\text{mean}$, and show power law decay or growth at late time.
Assuming the power-law dependence $H_n \propto v^{p_n}$,
we can obtain the powers $p_n$ by fitting the numerical data.
We summarize the resulting $p_n$ in Fig.~\ref{pOFm2}.
The numerical results strongly suggest for general $m$
\begin{equation}
H_n \propto v^{p_n}\,,
\qquad
p_n = n + 
1 - \Delta_m
~.
\label{power_H_massive}
\end{equation}
Using the results in section \ref{Sec:massiveAdS2},
we can see that this is the same scaling as appears for a massive scalar field in an $AdS_2$ background.

Note in Panel (b) of Fig.~\ref{mvar_H} that $p_1$ is positive for $(Mm)^2=1$ which confirms the existence of an instability in this case, for which the analytic argument above does not apply. The instability involves blow-up $\partial_r^2 \phi|_{r=M}$, in agreement with the prediction of (\ref{ngen}).

\begin{figure}[htbp]
\centering
\includegraphics[width=7.5cm, clip]{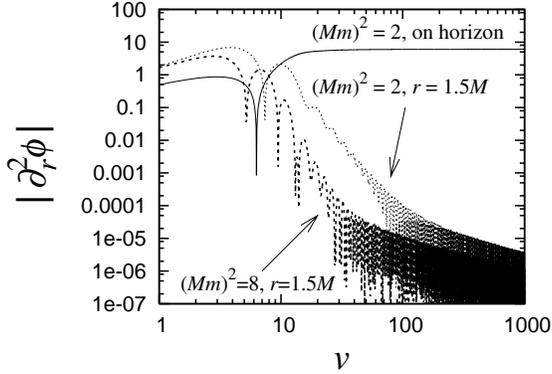}
\caption{$\left|\partial_r^2\phi\right|$ on the horizon for $(Mm)^2=2$
contrasted with that at $r=1.5M$ for $(Mm)^2=2,8$.
$\partial_r^2\phi\big|_{r=M}$ for $(Mm)^2=2$
converges to a constant at late time as suggested by the general argument,
while $\partial_r^2\phi$ off the horizon decays as $\propto v^{-5/6}$ irrespective of
the mass $m$.
}
\label{ddphi_r1.5}
\end{figure}

We close this section with some comments on the behaviour of derivatives of the field outside the horizon.
In Fig.~\ref{ddphi_r1.5}, we show $\partial_r^2 \phi$ on the horizon for $(Mm)^2=2$,
and also that quantity off the horizon for $(Mm)^2=2,8$.
As expected from the general argument, $\partial_r^2 \phi\big|_{r=M}$ converges into a constant 
at late time. On the other hand, $\partial_r^2 \phi$ outside the horizon 
shows damped oscillation whose amplitude decays as $v^{-5/6}$.
This behaviour appears universal for any scalar field mass $m$ and also order of the derivative, 
that is, the envelope of the damped oscillation of $\partial^k_r \phi$ is proportional to $v^{-5/6}$ for any $k$ and $m$. 
Conservation or blow up of $\partial^k_r \phi$ is a property that occurs only {\it on} the horizon
of an extreme black hole, and away from the horizon these quantities simply decay (as proved by Aretakis in the massless case).

\subsubsection{Zero conserved quantity}

In this section, 
we briefly comment on the cases $H_n[\psi]=0$ for the massive scalar field.
We focus on the case $n=1$ ($(Mm)^2 = 2$), but the results for general $n$ are analogous.

\begin{figure}[htbp]
\centering
\includegraphics[width=7.5cm, clip]{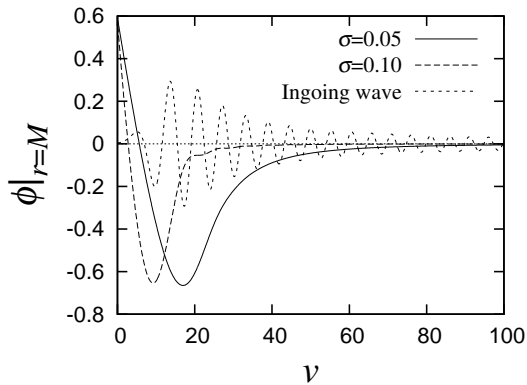}
\caption{
$\phi|_{r=M}(v)$ for $(Mm)^2=2$ with $H_1=0$.
We used the outgoing wave initial data with $\sigma = 0.05$ and $0.1$,
and also the ingoing wave initial data with $(\sigma,\mu)=(3.0, 10.0)$.
For the outgoing wave initial data, $\mu$ is fixed as
$\mu = - \sigma \left(
\sigma + \sqrt{\sigma^2 + 4M^2}
\right)/(2M)$ by the requirement of $H_1=0$.
}
\label{phiH_H0_massive}
\end{figure}

First, we show the time dependence of $\phi|_{r=M}$ in Fig.~\ref{phiH_H0_massive}. 
To set $H_1=0$, we used 
the outgoing wave initial data with 
$\mu = -\sigma\left(\sigma+\sqrt{\sigma^2+4M^2}\right)/(2M)$.
The behaviour for the outgoing wave initial data is similar to the $H_n \neq 0$ case 
studied above in that $\phi^\text{mean}$  and $\phi^\text{osci}$ are non-negligible at $r=M$. 
On the other hand, $\phi|_{r=M}(v)$ for the ingoing wave initial data is dominated by the oscillatory component with negligible contribution from $\phi^\text{mean}$.

\begin{figure}[htbp]
\centering
\subfigure[$\phi|_{r=M}^\text{mean}$]{
\includegraphics[width=7.5cm, clip]{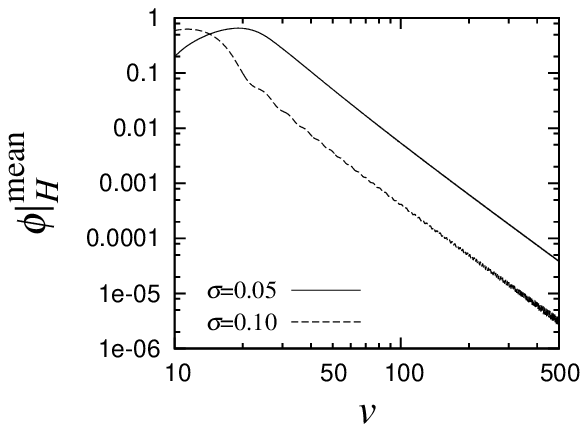}
}
\subfigure[$A|^\text{osci}_{r=M}$]{
\includegraphics[width=7.5cm, clip]{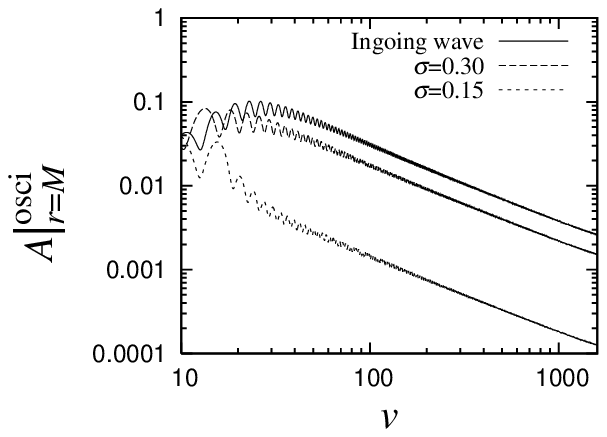}
}
\caption{
Panel~(a): $\phi|_{r=M}^\text{mean}(v)$ of $(Mm)^2=2$ and $H_1=0$ for
the outgoing wave initial data with $\sigma = 0.05$ and $0.1$.
Fitting the curves with the function $Cv^a$, we find the power of the decay to be
$a=-3.01$ and $-2.99$ for each initial data.
We do not show the result for the ingoing wave initial data since 
$\phi|_{r=M}^\text{mean}$ is almost vanishing compared to $\phi|_{r=M}^\text{osci}$ for it.
Panel~(b): Amplitude of the damped oscillation of $\phi|_{r=M}^\text{osci}(v)$, $A|^\text{osci}_{r=M}(v)$,
for 
the ingoing wave initial data with $(\sigma,\mu)=(3.0,10.0)$
and 
outgoing wave initial data with $\sigma = 0.30, 0.15$.
Fitting the data, we find the decay $-0.85$, $-0.84$ and $-0.84$.
These results suggests that $A|_{r=M}^\text{osci}(v)$ becomes proportional to $v^{-5/6}$ at 
late time, which is the same power law as in the $H_1\neq 0$ case studied above.
}
\label{H0_massive}
\end{figure}

We summarize the behaviour of the mean and oscillatory components of $\phi|_{r=M}$ 
in Fig.~\ref{H0_massive}.
First, we find from Panel~(a) that the mean component shows the power-law 
decay $\phi|_{r=M}^\text{mean}\propto v^{-3}$, which is one power faster than the 
$H_1\neq 0$ case studied above. This property is similar to that of the massless 
case with $l=1$, and we can show that $\partial_r^4\phi|_{r=M}$ diverges linearly in time (as it did in the massless case). Hence, just as in the massless case, vanishing $H_1$ requires one more $r$-derivative to see the instability.

Next, we find from Panel~(b) that the amplitude of the oscillatory component 
shows the power-law decay $A|^\text{osci}_{r=M}(v) \propto v^{-5/6}$ at late time.
This is the same power law as the $H_1\neq 0$ case shown in Fig.~\ref{norm_m2},
and thus the time dependence of $A_\text{osci}$ seems to be insensitive to the value
of the conserved quantity.

To sum up, the behaviour of the component $\phi|^\text{mean}_{r=M}$ of a 
massive scalar field with $(Mm)^2=2$ can be described by Eq.~(\ref{latel1}), 
which was proposed for the massless scalar field with $l=1$.
For general value of the scalar field mass $m$, our results for $\phi|^\text{mean}_{r=M}$ are consistent with 
a power-law decay $v^{-\Delta_m}$ for generic initial data,
or $v^{-\Delta_m - 1}$ for initial data with $H_1=0$, where $\Delta_m$ is defined below Eq.~(\ref{power_H_massive}).
On the other hand, the amplitude of the damped oscillation of $\phi|^\text{osci}_{r=M}$ or the derivative
$\partial_r^k \phi$ {\it outside} the horizon, appears to be proportional, at late time, to $v^{-5/6}$ irrespective of $m$ and also of order of the derivative.

\section{Gravitational and electromagnetic perturbations}

\label{gravem}

In this section we show that an analogue of Aretakis' scalar field instability occurs for the coupled linearised gravitational and electromagnetic perturbations of the extreme RN solution. To demonstrate this we will employ past work on the linearized perturbations of extreme RN. Several formalisms were developed for studying this problem, see~\cite{Zerilli:1974ai, Moncrief:1974gw, Moncrief:1974ng, Moncrief:1975sb, Chitre:1976bb, Lee, Chandrasekhar:1985kt} and also~\cite{bicak2} which relates the various methods. These works established the {\it mode} stability of the RN black hole (including the extreme case). 

We will use Moncrief's gauge invariant formalism~\cite{Moncrief:1974gw, Moncrief:1974ng, Moncrief:1975sb} due to the remarkable simplicity of the resulting perturbation equations. This approach is based on the Hamiltonian formulation of the Einstein-Maxwell equations, where the perturbation variables are certain gauge invariant combinations (under both infinitesimal diffeomorphisms and electromagnetic gauge transformations) of components of the metric and gauge field.  We will first briefly explain Moncrief's formalism and variables, then we present conservation laws for these variables and associated instability results.

\subsection{Moncrief's gauge invariant formalism}

Moncrief simplified the perturbation equations for the RN solution by employing a reduction method based on the Hamiltonian formulation of Einstein-Maxwell theory of ADM. 

First recall that in the Hamiltonian formulation one chooses a preferred time function $t$ and decomposes the spacetime fields on hypersurfaces $\Sigma_t$ of constant $t$, with coordinates $x^i$ for $i=1,2,3$, as follows. The metric $g_{\mu\nu}$ is written in terms of the lapse $N=-(g^{tt})^{1/2}$, shift $N_i=g_{ti}$ and the induced $3$-metric $g_{ij}$ on $\Sigma_t$, whereas the gauge field $A_\mu$ is decomposed in terms of the scalar potential $A_t$ and the spatial vector potential $A_i$.  The canonically conjugate momenta to $g_{ij}$ and $A_i$ are the tensor  densities $\pi^{ij}=\sqrt{g}(K^{ij}-Kg^{ij})$ and $E^i=N \sqrt{g}F^{ti}$ respectively, where $K_{ij}$ is the extrinsic curvature on $\Sigma_t$. Variation of the action functional with respect to these canonically conjugate pairs leads to dynamical evolution equations, whereas variation with respect to $N, N_i,A_t$ leads to the constraint equations $\mathcal{H}=\mathcal{H}_i= \mathcal{E}=0$.

We are now ready to give a brief review Moncrief's reduction method, see~\cite{Moncrief:1975sb} for more details.  The perturbation equations can be derived by taking the second variation of the action functional evaluated on an exact solution. The canonical variables for the linearised ADM equations are simply the first order perturbations of the canonical variables $(g_{ij}, \pi^{ij})$ and $(A_i, E^i)$, which we will denote by $(\delta g_{ij}, \delta \pi^{ij})$ and $(\delta A_i, \delta E^i)$. Similarly, the constraint equations for the perturbed problem are the linearised constraints $\delta \mathcal{H}=\delta\mathcal{H}^i =\delta \mathcal{E}=0$ of the exact problem; these must be conserved in time as a consequence of the evolution equations for the canonical variables. There are two crucial facts about the linearised constraint functions $\delta \mathcal{H},\delta\mathcal{H}^i,\delta \mathcal{E}$: they generate infinitesimal diffeomorphisms and electromagnetic gauge transformations, and mutually commute under the Poisson bracket. These properties allow one to perform a canonical transformation to new canonical variables such that $\delta \mathcal{H},\delta\mathcal{H}^i,\delta \mathcal{E}$ are a subset of the momenta and their conjugate variables are cyclic  and gauge-dependent.  Hence the remaining canonical pairs, which by construction now will commute with $\delta \mathcal{H},\delta\mathcal{H}^i,\delta \mathcal{E}$, must be gauge-invariant.  In terms of these new canonical variables the problem greatly simplifies: the evolution equations for the gauge-invariant pairs decouple from all gauge dependent quantities and the constrains can be eliminated.

Moncrief applied the above general method to perturbations of the RN solution: naturally the time function $t$ was chosen to correspond to the static time coordinate. The background data is thus (we don't yet assume extremality)
\bea
N^2 &=&F(r) \equiv 1-\frac{2M}{r}+\frac{Q^2}{r^2} \; , \qquad g_{ij}dx^idx^j = \frac{dr^2}{F(r)} +r^2 d\Omega^2, \label{ADMRN1}\\  
N_i &=&0 , \qquad \pi^{ij}=0, \qquad  A_i = 0, \qquad E^r =  2Q \sin \theta  \; . \label{ADMRN2}
\eea
Spherical symmetry of the background solution allows one to further simplify the problem. One expands all perturbation variables in terms of spherical harmonics $Y_{lm}$ which then leads to decoupled equations for each value of $l= 0, 1,2, \dots$. We now describe the basis used to do this, which is essentially that used by  Regge and Wheeler.

There are three classes of perturbation depending on whether they transform as a scalar, vector or tensor on $S^2$. A basis for each of these (written covariantly) is given by the following harmonics:
\bea
&&\text{Scalar ($l\geq 0$)}:  \quad Y=Y_{lm}  \qquad \text{where} \qquad -\hat{\nabla}^2 Y = l(l+1) Y \label{S2scalar} \\
&&\text{Vector ($l\geq 1$)}:  \quad  \hat{e}_a= -\hat{\epsilon}_{a}^{~b} \partial_b Y,  \qquad  \hat{f}_a= \partial_aY \label{S2vector}\\
&& \text{Tensor ($l\geq 2$)}: \quad  \hat{e}_{ab}=\hat{\epsilon}_{(a}^{~~c} \hat{\nabla}_{b)} \hat{\nabla}_{c} Y, \qquad  \hat{f}_{ab}= \hat{\nabla}_a \hat{\nabla}_b Y,  \qquad  \hat{g}_{ab}= Y(d\Omega^2)_{ab}, \quad  \label{S2tensor}
\eea
where $d\Omega^2, \hat{\nabla}_a, \hat{\epsilon}_{ab}$ are the round metric, the associated connection and volume form on the unit $S^2$ and $a,b, \dots$ are $S^2$ indices. As always the $Y_{lm}$ are orthonormal with respect to the unit $S^2$ measure; from the properties of $Y$ one may deduce corresponding properties for the vector and tensor harmonics. These harmonics all have a well defined parity under the discrete isometry which reverses the orientation. In particular $Y, \hat{f}_a, \hat{f}_{ab}, \hat{g}_{ab}$ have parity $(-1)^l$, whereas $\hat{e}_a, \hat{e}_{ab}$ have parity $(-1)^{l+1}$. Traditionally these are referred to as {\it even} or {\it odd} parity respectively; for each $l$ the perturbation equations for the two types of parity therefore decouple.  We note that since we are in four dimensions perturbations of the tensor components of the Maxwell two form can be expanded in the basis $Y \hat{\epsilon}_{ab}$, which is an odd parity harmonic.

By Birkhoff's theorem any $l=0$ perturbation must correspond to the RN family of solutions.  Perturbations which correspond to a slowly rotating Kerr-Newman black hole are contained in the $l=1$ odd type, as discussed below. Hence any instability must either be an $l \geq 2$ perturbation or a  non-stationary $l=1$  perturbation.

Moncrief showed~\cite{Moncrief:1974gw, Moncrief:1974ng, Moncrief:1975sb} that each type of perturbation can be fully described by either one gauge invariant variable ($l=1$) or a pair of gauge invariant variables ($l >1$). We describe these below and more fully in Appendix \ref{app:moncrief}. Remarkably, each of these variables satisfies a decoupled wave equation of the general form:
\be
\label{genformstatic}
-\frac{r^2}{\Delta} \partial_t^2\psi + \partial_r(r^{-2}\Delta \partial_r \psi) = r^{-2}W(r) \psi
\ee
where $\Delta= r^2F$, for some explicitly known function $W(r)$. However, one shortcoming of Moncrief's approach is that since the various $\psi$ are defined using static coordinates, it is not obvious if they are smooth on the horizon (since those coordinates breakdown there). In Appendix \ref{app:moncrief} we show that in fact all the relevant gauge invariant variables $\psi$ are indeed smooth on the future horizon $\mathcal{H}^+$. Hence we may convert to EF coordinates resulting in the master equation
\be
\label{genform}
2 \partial_v \partial_r \psi + \partial_r(r^{-2}\Delta \partial_r \psi) = r^{-2}W(r) \psi  \; .
\ee
All terms in this equation are smooth on $\mathcal{H}^+$ which is at the largest root $r=r_+$ of $\Delta$.

The odd parity perturbations are simplest, so let us describe the gauge invariant variables in this case (see Appendix \ref{app:moncrief} for more detail on their construction).  

General $l \geq 1$ odd perturbations of the gauge field must be of the form $\delta A_a =A\hat{e}_a$ for some function $A$. It turns out $\pi_f=A$ is one of Moncrief's gauge-invariant canonical momentum variables. This variable can be defined invariantly as follows:
\be
\pi_f \equiv  \frac{1}{l(l+1)}\int_{S^2(v,r)} d\Omega \; (\hat{e}^a)^* \delta A_a = \frac{1}{l(l+1)} \int_{S^2(v,r)} Y^*_{lm} \delta F   \label{pif}
\ee
where $S^2(v,r)$ a sphere of constant $(v,r)$ in EF coordinates and the second equality follows from integration by parts. Under a gauge transformation $\delta F = L_\xi F= d i_\xi F$ where $\xi^\mu$ is a  vector generating infinitesimal diffeomorphisms. For odd perturbations $\xi_\mu dx^\mu = C(v,r) \hat{e}_a dx^a$ and it is readily checked that this implies $\delta F_{ab}=0$. Hence $\pi_f$ is indeed gauge invariant and the above definition shows it is also smooth on the horizon.

The relevant variable to describe $l \geq 1$ odd perturbations of the metric turns out to be $p_1$, which is the conjugate momentum to the metric perturbation $\delta g_{ra} = h_1 \hat{e}_a$.  In Appendix \ref{app:moncrief} we give details of general odd perturbations of the metric and show that\be
p_1  = \int_{S^2(v,r)}  d\Omega\;  (\hat{e}^a)^* \left( \partial_v \delta g_{r a} -\partial_r \delta g_{va} + \frac{2 \delta g_{va}}{r}  \right)  \label{pi1} \; ,
 \ee
where here $\delta g_{\mu\nu}$ is written in EF coordinates, which explicitly shows $p_1$ is a smooth quantity on $\mathcal{H}^+$. It is easily checked this is gauge-invariant directly. For odd perturbations an infinitesimal diffeomorphism must be of the form $\delta g_{\mu\nu} = \nabla_{( \mu} \xi_{\nu)}$  where again $\xi_\mu dx^\mu = C(v,r) \hat{e}_a dx^a$. Hence $\delta g_{va}= \partial_v C$ and $\delta g_{ra} = \partial_r C - 2C/r$;  it follows that $\delta \pi_1=0$ as claimed.
 
Moncrief showed that $l>1$ odd perturbations reduce to two decoupled wave equations  (\ref{genform}) for $\psi= P_\pm$ where $P_\pm$ are certain constant linear combinations of $\pi_f$ and 
\be
\pi_g \equiv r p_1 - \frac{2Q l(l+1)}{r} \pi_f  \label{pig}  \; .
\ee
From above we deduce that $\pi_f$, $\pi_g$ and hence $P_\pm$ are indeed smooth functions on the horizon.

For $l=1$ odd perturbations Moncrief showed that the quantity $\delta a\equiv - \tfrac{1}{6M}(\tfrac{1}{2}r^2 p_1- 2Q \pi_f)$, which from the above is gauge invariant and smooth on $\mathcal{H}^+$, is in fact a constant. In this case the perturbation equations reduce to a single wave equation of the general form (\ref{genform}) with
\be
\label{l1odd}
\psi = P_f \equiv  \pi_f - \frac{2Q \delta a}{r} , \qquad \qquad  W = 2\left( 1+ \frac{2Q^2}{r^2} \right)  \; .
\ee
From the above we deduce that $P_f$ is a gauge invariant quantity which is smooth on the horizon (since both $\pi_f$ and $\delta a$ are). Note that $P_f \equiv 0$ corresponds to a perturbation to a slowly rotating Kerr-Newman black hole with rotation parameter $\delta a$ (explaining the notation); indeed this is the most general regular stationary solution in this case~\cite{bicak2}.
 
The even parity perturbations are more complicated to describe. In Appendix \ref{app:moncrief} we explain their definition and show that the gauge invariant quantities appearing in Moncrief's final decoupled equations (\ref{genform}) are also all smooth on the horizon. We establish this by showing how to express them as linear combinations of the metric and field strength perturbations in EF coordinates.

\subsection{Conservation laws and blow up on the horizon}

As explained in the previous section, Moncrief's perturbation equations can all be reduced to the form (\ref{genform}) for some gauge invariant function $\psi$ which is regular on $\mathcal{H}^+$. We will now restrict to the extreme RN solution $Q=M$ so $\Delta =(r -M)^2$ and $r_+=M$. Using this general form we now derive necessary and sufficient conditions for the existence of certain conserved quantities on the horizon. As we will show below, remarkably, these conditions are satisfied for all types of $l \geq 1$ perturbation. Hence coupled gravitational and electromagnetic perturbations of an extreme RN solution possess a tower of conservation laws on the horizon. In turn, these conserved quantities lead to blow up results and hence instabilities for all types of perturbation.

Let $f(r)$ be a smooth function which is non-vanishing on the horizon and without loss of generality set $f|_{r=M} =1$. Multiply (\ref{genform}) by $r^2f(r)$ and differentiate $p \geq 2$ times with respect to $r$ and then set $r=M$. This gives an equation of the form
\be
\label{pdiff}
2 \partial_v[ \partial_r^p( r^2 f \partial_r \psi) ]|_{r=M}+ \sum_{k=0}^{p} a_k \partial_r^{p-k}\psi|_{r=M} = 0  \; ,
\ee
where in particular
\be
\label{a0}
a_0 = p(p+1)-W|_{r=M} . \;
\ee
Hence the quantity in square brackets in (\ref{pdiff}) is conserved on the horizon if and only if $a_k=0$ for all $0 \leq k \leq p$.  In turn, this is equivalent to the following set of equations:
\bea
W|_{r=M} &=& p(p+1) \label{WH}\\ 
\label{eqf}
f^{(k)}|_{r=M} &=& - \frac{1}{2p+1-k} \left[ \frac{1}{k} \sum_{q=1}^k \left( \begin{array}{c} k \\ q \end{array} \right) f^{(k-q)} W^{(q)} + 2(p-k) (r^{-1} f)^{(k-1)} \right]_{r=M} \label{fder} 
\eea
where $1\leq k \leq p$.  For each $k \geq 1$ equation (\ref{fder}) determines $f^{(k)}|_{r=M}$ in terms of the lower order derivatives $f^{(k-1)}|_{r=M}, \dots, f|_{r=M}$. It follows that  for all $1 \leq k \leq p$ the $f^{(k)}|_{r=M}$ are determined (recall we have fixed $f|_{r=M}=1$). Hence by choosing $f$ to be any function whose derivatives satisfy these constraints we deduce that
\be
\label{Hp}
H_p[\psi] \equiv\frac{1}{M^2} [ \partial_r^p( r^2 f \partial_r \psi) ]|_{r=M}
\ee
is conserved along $\mathcal{H}^+$ if and only if (\ref{WH}) is satisfied. Notice that $H_p$ depends only on the first $p$ derivatives of $f$ at $r=M$ and is therefore independent of any specific choice of $f$.

This conservation law implies generic non-decay at late time on ${\cal H}^+$ of a certain linear combination of $\psi$ and its first $(p+1)$ $r$-derivatives. Since we expect all derivatives of the field to decay {\it outside}\footnote{
 It would be surprising (and very interesting) if this expectation were false because then metric-Maxwell perturbations would exhibit  behaviour that is worse than that of scalar field perturbations. The case $l=1$ is an exception for which the exterior solution might settle down to a perturbation within the Kerr-Newman family.} ${\cal H}^+$, it follows that certain quantities with one more $r$-derivative will blow up at late time on ${\cal H}^+$. 

To see this explicitly, suppose we have a conserved quantity $H_p[\psi]$ for some $p \geq 2$ as above. Now consider (\ref{pdiff}) with $p\to p+1$ so
\be
\label{p1diff}
2 \partial_v[ \partial_r^{p+1}( r^2 f \partial_r \psi) ]|_{r=M}+ 2(p+1) \partial_r^{p+1}\psi|_{r=M}+ \sum_{k=1}^{p+1} a_k \partial_r^{p+1-k}\psi|_{r=M} = 0 
\ee
where we have used (\ref{WH}) to simplify the coefficient of $\partial_r^{p+1}\psi|_{r=M}$ which in this case is non vanishing.
Now (motivated by the results for the scalar field) suppose that $\partial_r^k\psi|_{r=M} \to 0$ as $v\to \infty$ for $0 \leq k \leq p$.  Then (\ref{Hp}) implies $\partial_r^{p+1} \psi|_{r=M} \to H_p[\psi]$. Hence (\ref{p1diff}) implies
\be
\partial_v [ \partial_r^{p+1}( r^2 f \partial_r \psi) ]|_{r=M} \sim - (p+1)H_p
\ee
as $v \to \infty$. Therefore if the constant $H_p \neq 0$, as will be the case for generic initial data, we deduce the linear blow up
\be
\partial_r^{p+2} \psi|_{r=M} \sim - \frac{(p+1)H_p v}{M^2}
\ee
as $v\to \infty$. Iterating this argument shows
\be
\partial_r^{p+n+1} \psi|_{r=M} \sim - c_{p,n} H_p v^n
\ee
for some non-zero constants $c_{p,n}$ and all $n\geq 1$.

Let us now apply these results to Moncrief's perturbation equations. As we show below, remarkably, for all types of perturbation there exists some integer $p\geq 2$ such that (\ref{WH}) is satisfied, as summarised in table \ref{table}. 
\begin{table}[!h]

\centering
\begin{tabular}{ | c | c | c | c | c| }
\hline
 & $l=1$ odd & $l>1$ odd  & $l=1$ even & $l>1$ even\\
\hline
 $\psi$  & $P_f$ & $P_{\pm}$  & $H$ & $R_{\pm}$ \\
 \hline
 $W|_{r=M}$ & $6$ & $ l(l+1)+1 \pm (2l+1)$ & $6$ & $l(l+1)+1 \pm (2l+1)$ \\
 \hline
 $p$ & $2$ & $l \pm 1$ & $2$ & $l \pm 1$ \\
 \hline
 \end{tabular}
 \caption{Conserved quantities $H_p[\psi]$ for Moncrief's perturbations.}
 \label{table}
\end{table} \\
Hence for all types of perturbation there exists a conserved quantity (\ref{Hp}) and if this quantity is non-zero, there exists an associated blow up result for the corresponding transverse derivatives. Note that the smallest value of $p$, i.e., the fewest transverse derivatives, is $p=1$ which occurs for $l=2$.

Consider the $l=1$ odd perturbation equation (\ref{l1odd}) for which $\psi= P_f$ and $W|_{r=M}=6$. In this case there exists a conserved quantity for $p=2$: explicitly one can use (\ref{fder}) to get $f'|_{r=M}=3/2$ and $f''|_{r=M}=0$ and deduce that
\be
H_2[P_f] = \left( \partial_r^3 P_f + \frac{7}{M} \partial_r^2 P_f + \frac{8}{M^2} \partial_r P_f \right)_{r=M}  \; .
\ee
For a generic initial perturbation, $H_2[P_f] \neq 0$ and hence $\partial_r P_f, \partial_r^2 P_f$ and $\partial_r^3 P_f$ cannot all decay at late time on ${\cal H}^+$. In view of the results for a scalar field, the most likely behaviour is that $P_f, \partial_r P_f, \partial_r^2 P_f$ all decay along the horizon and hence $\partial_r^3 P_f$ does not decay. In this case, from the above argument, $\partial_r^4 P_f$ will blow up linearly in $v$ on ${\cal H}^+$. Hence we must have an instability.  This is readily translated to a statement about explicit components of the Maxwell field: if we restrict to perturbations for which $\delta a =0$ (e.g.\ by demanding vanishing angular momentum perturbation), then we deduce the best possible generic behaviour is that the $S^2$ components $\delta F_{ab}, \partial_r \delta F_{ab}, \partial_r^2 \delta F_{ab}$ all decay, $\partial_r^3 \delta F_{ab}$ does not decay and $\partial_r^4 \delta F_{ab}$ must blow up linearly in $v$ along $\mathcal{H}^+$.

Now consider $l>1$ odd perturbations.  In this case the variables $\psi=P_\pm$ are related to the gauge-invariant variables $\pi_f, \pi_g$ defined in equations (\ref{pif}), (\ref{pig}) and (\ref{pi1}), by 
\bea
\pi_f &=& \frac{1}{\sqrt{2(2l+1)(l-1)(l+2)}} \left[ \sqrt{2l+4} \;P_+ - \sqrt{2l-2} \; P_- \right]  \label{Ppm1}\\
\pi_g &=& \frac{l(l+1)}{\sqrt{2(2l+1)}} \left[ \sqrt{2l-2} \;P_+ + \sqrt{2l+4} \; P_-\right]  \; , \label{Ppm2}
\eea
whereas the associated functions appearing in the wave equation (\ref{genform}) are
\be
W_\pm =  l(l+1)-\frac{3M}{r}+\frac{4M^2}{r^2} \pm \frac{(2l+1)M}{r}  \; .
\ee
Therefore $W_{\pm}|_{r=M}= l(l+1)+1 \pm (2l+1)$ and hence $W_+|_{r=M} = (l+1)(l+2)$ and $W_-=l(l-1)$. We deduce that for $P_+$ there exists a conserved quantity for $p=l+1$ and for $P_-$ a conserved quantity for $p=l-1$.\footnote{
The recurrence relation~(\ref{eqf}) can be solved explicitly in this case. For $P_+$
$$
f^{(k)}|_{r=M} = 
-\frac{3 k!(-1)^k(l-k+1)(2l-k+2)}{M^k l(l+1)(2l+1)}
$$
and for $P_-$
$$f^{(k)}|_{r=M}= 
\frac{
(-1)^k k! (k+1) (l-k-1)
}{
M^k(l-1)  
} \; .
$$
}

For definiteness consider the $l=2$ odd case so then we have conserved quantities:
\bea
H_1[P_-] &=& \left( \partial_r^2 P_-+ \frac{2}{M} \partial_r P_- \right)_{r=M}\\
H_3[P_+] &=& \left( \partial_r^4 P_++... \right)_{r=M}
\eea
where the ellipsis denotes lower order $r$-derivatives of $P_+$, whose explicit coefficients are unimportant.%

For a generic initial perturbation these constants will be non-zero (and independent). Hence $\partial_r P_-$ and $\partial_r^2 P_-$ cannot both decay at late time on ${\cal H}^+$. One can relate $P_-$ directly to the metric and Maxwell field perturbation by writing it as a linear combination of $\pi_f$ and $\pi_g$ using (\ref{Ppm1}) and (\ref{Ppm2}). We can then use (\ref{pig}) to write $P_-$ in terms of $\pi_f$ and $p_1$. Recall $p_1$ is the gauge invariant combination of metric perturbation components (\ref{pi1}) and $\pi_f$ is the gauge invariant Maxwell field perturbation (\ref{pif}). Hence we deduce that a certain gauge-invariant linear combination of the first two $r$-derivatives of these quantities generically does not decay at late time on ${\cal H}^+$.\footnote{
We emphasize that this non-decay cannot be because the perturbation is settling down to a time-independent perturbation corresponding to a variation of parameters within the Kerr-Newman family because $l=2$ here.}

As discussed above, it seems very likely that, for $l=2$, the gauge invariant quantities and all of their derivatives will decay at late time {\it outside} ${\cal H}^+$. Hence, since we have non-decay of a certain quantity {\it on} ${\cal H}^+$, there will be a quantity with one more $r$-derivative which blows up at late time on ${\cal H}^+$. The most likely scenario is that $P_-$ and $\partial_r P_-$ will decay at late time along the horizon, in which case $\partial_r^2 P_-$ cannot decay and $\partial_r^3 P_-$  will blow up at late time on ${\cal H}^+$. Similarly, if $P_+, \partial_r P_+, \partial_r^2 P_+, \partial_r^3 P_+$ all decay along $\mathcal{H}^+$, then $\partial_r^4 P_+$ cannot decay and $\partial_r^5 P_+$ will blow up at late time on ${\cal H}^+$. It then follows that $\pi_f,\pi_g, \partial_r \pi_f, \partial_r \pi_g$ decay, whereas 
\bea
&&\partial_r^2 \pi_f|_{r=M} \sim -\frac{H_1[P_-]}{2\sqrt{5}}   \qquad \qquad \partial_r^3 \pi_f|_{r=M} \sim \frac{H_1[P_-]}{\sqrt{5} M^2} v \\
 &&\partial_r^2 \pi_g|_{r=M} \sim  \frac{12 H_1[P_-]}{\sqrt{5} } \qquad \qquad \partial_r^3 \pi_g|_{r=M} \sim - \frac{24 H_1[P_-]}{\sqrt{5} M^2} v
\eea
in which case from equation (\ref{pig}) it follows that $p_1, \partial_r p_1$ decay and
\be
\partial_r^2 p_1|_{r=M} \sim \frac{6 H_1[P_-]}{\sqrt{5} M} \qquad \partial_r^3 p_1|_{r=M} \sim -\frac{12 H_1[P_-]}{\sqrt{5} M^3}   v    \; ,
\ee
as $ v \to \infty$.
Hence there is an $l=2$ instability where transverse derivatives of these gauge invariant perturbations blow up linearly on the horizon.

Similar statements for $l=1$ even and $l>1$ even perturbations can be deduced from table \ref{table} and Appendix \ref{app:moncrief}.  The associated functions $W$ in (\ref{genform}) for each variable $\psi$ are easily read off from~\cite{Moncrief:1974ng, Moncrief:1975sb} and~\cite{bicak2};  in table \ref{table} we provide $W|_{r=M}$ which as shown above is enough to determine the existence of a tower of conservation laws and associated instabilities on the horizon.

\subsection*{Acknowledgments}

We are grateful to Gary Gibbons for telling us about the conformal isometry of extreme RN, and for suggesting that this may relate the Aretakis constants to the NP constants. We are also grateful to Mihalis Dafermos, Shamit Kachru and Helvi Witek for discussions. 
NT acknowledge hospitality 
during the YITP Long-term workshop YITP-T-12-03 on ``Gravity and Cosmology 2012''
at Yukawa Institute for Theoretical Physics at Kyoto University 
where part of this work was completed.
A part of numerical computation in this work was carried out at the Yukawa Institute Computer Facility.
JL is supported by an EPSRC Career Acceleration Fellowship. 
KM is supported by JSPS Grant-in-Aid for
Scientific Research No.24$\cdot$2337.
HSR is supported by a Royal Society University Research Fellowship and by European Research Council grant no. ERC-2011-StG 279363-HiDGR.
N.T.\ is supported in part by the DOE Grant DE-FG03-91ER40674.

\appendix
\section{Numerical method}

\subsection{Algorithm}
\label{ALG}
Here, we explain our numerical method to solve Eq.~(\ref{waveUv}).
We take the domain of the numerical calculation as
$U\in [-0.5,0]$ and $v\in [0,2000]$.
We discretize coordinates $U$ and $v$ as $(U_i,v_j)$ 
($i=0,1,2,\cdots,N$, $j=0,1,2\cdots M$) where
$U_0=-0.5$, $U_N=0$, $v_0=0$ and $v_M=2000$.
Denoting $\phi_{i,j}=\phi(U_i,v_j)$, $V_{i,j}=V(U_i,v_j)$,
$\delta U_i=U_{i+1}-U_i$ and $\delta v_j=v_{j+1}-v_j$,
we obtain the discretized equation for Eq.~(\ref{waveUv}) as
\begin{equation}
\phi_{i+1,j+1}=\phi_{i,j+1}+\phi_{i+1,j}-\phi_{i,j}
-\frac{\delta U_i \delta v_j}{4}(V\phi)_{i+1/2,j+1/2}
+\mathcal{O}(\delta U_i^3\delta v_j, \delta U_i \delta v_j^3)\ ,
\end{equation}
where 
$(V\phi)_{i+1/2,j+1/2}=(V_{i+1,j+1}\phi_{i+1,j+1}+V_{i+1,j}\phi_{i+1,j}+V_{i,j+1}\phi_{i,j+1}+V_{i,j}\phi_{i,j})/4$.
Solving above equation for $\phi_{i+1,j+1}$, we have
\begin{multline}
\label{diseq}
 \phi_{i+1,j+1}=\left(1+\frac{\delta U_i \delta v_j}{16}V_{i+1,j+1}\right)^{-1}
\left(\phi_{i,j+1}+\phi_{i+1,j}-\phi_{i,j}
-\frac{\delta U_i \delta v_j}{4}(V\phi)'_{i+1/2,j+1/2}\right)\\
+\mathcal{O}(\delta U_i^3\delta v_j, \delta U_i \delta v_j^3)\ ,
\end{multline}
where 
$(V\phi)'_{i+1/2,j+1/2}=(V_{i+1,j}\phi_{i+1,j}+V_{i,j+1}\phi_{i,j+1}+V_{i,j}\phi_{i,j})/4$.
Note that $\phi_{i,0}$ and $\phi_{0,j}$ are given by
the initial condition~(\ref{out}) or (\ref{ing}).
Thus, using Eq.~(\ref{diseq}), 
we can determine the $\{\phi_{i,j}\}$ in the whole domain recursively.

\begin{figure}
\begin{center}
\includegraphics[scale=0.5]{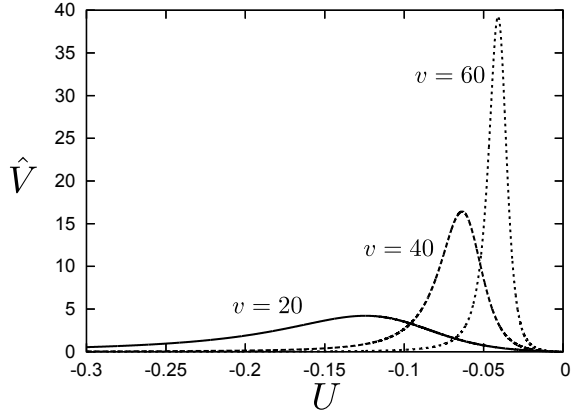}
\end{center}
\caption{
The potential $\hat{V}(U,v)$ for $m=0$ and $l=0$.
We take fixed $v$ slices: $v=20,40,60$. 
We can see that the potential approaches $U=0$ and 
tends to be sharp as $v$ increases.
}
 \label{Vpot}
\end{figure}

Now, we consider how we should take the numerical grid, $(U_i,v_j)$.
In Fig.~\ref{Vpot}, 
we depict the potential $\hat{V}(U,v)$ for fixed $v$ slices.
We can see that the potential approaches $U=0$ and 
tends to be sharp as $v$ increases.
This implies that 
we need smaller grid size as $U$ approaches zero.
Thus, we take a multigrid generated by following algorithm: 
$U_i=U_{i+1}-c_i h$ where $U_N=0$ and $c_i=1,4,16,64,256$ for 
$0\leq i < N'$,
$N'\leq i <2N'$,
$2N'\leq i <3N'$,
$3N'\leq i <4N'$ and
$4N'\leq i <5N'$, respectively.
Here, we define $N'=N/5$ and $h=0.5/\sum_{i=0}^{N-1}c_i$.
For this choice of $h$, we have $U_0=-0.5$.
In our calculation, we take the grid number as $N=4\times 10^4$.
By this choice of numerical grid,
we can resolve the potential $\hat{V}(U,v)$ within 
$v\lesssim 2000$.
For $v$-direction, we take uniform grid $\delta v_j=0.1$.

\subsection{Evaluating transverse derivatives on the horizon}
\label{ETD}
In sections \ref{Nummassless} and \ref{Nummassisve}, 
we evaluate $\partial_r^n \phi|_{r=M}$ ($n=1,2,3,4$) from the numerical
solution $\phi(U,v)$. The numerical differentiations tend to lose the
accuracy as $n$ increases.
Thus, we use a trick to evaluate
the differentiations. In the $(v,r)$-coordinates, the wave
equation~(\ref{waveUv}) can be written as
\begin{equation}
\label{wavevr}
 2\partial_v\partial_r\phi+F(r)\partial_r^2\phi + F'(r)\partial_r\phi
 -V(r)\phi=0\ ,
\quad
V(r)=\frac{F'(r)}{r}+\frac{l(l+1)}{r^2}+m^2\ .
\end{equation}
where $'\equiv d/dr$. From the equation, we obtain recursion equations
for $\partial_r^n \phi|_{r=M}$ $(n=1,2,\cdots)$ as
\begin{align}
&2\partial_v \partial_r\phi|_{r=M}=[V \phi]_{r=M}\ ,
\label{1st}\\
&2\partial_v \partial_r^2\phi|_{r=M}
=\left[-\frac{2}{M^2}\partial_r\phi+V\partial_r\phi+V'\phi\right]_{r=M}\
 ,
\label{2nd}\\
&2\partial_v \partial_r^3\phi|_{r=M}
=\left[-\frac{6}{M^2}\partial_r^2\phi
+\frac{12}{M^3}\partial_r\phi
+V\partial_r^2\phi+2V'\partial_r\phi+V''\phi
\right]_{r=M}\ ,\\
&2\partial_v \partial_r^4\phi|_{r=M}
=\bigg[
-\frac{12}{M^2}\partial_r^3\phi
+\frac{48}{M^3}\partial_r^2\phi
-\frac{72}{M^4}\partial_r\phi \notag\\
&\hspace{5cm}
+V\partial_r^3\phi+3V'\partial_r^2\phi+3V''\partial_r\phi
+V'''\phi
\bigg]_{r=M}\ .
\end{align}
These equations are obtained by operating $\partial_r^m$ 
and setting $r=M$ in Eq.~(\ref{wavevr}) where $m$ is taken to be $m=0,1,2,3$.
Once we know $\phi|_{r=M}$ from the numerical calculation, 
we can determine $\partial_r\phi|_{r=M}$ by integrating Eq.~(\ref{1st})
with respect to $v$.
From the $\phi|_{r=M}$ and $\partial_r\phi|_{r=M}$, 
we can also determine $\partial_r^2\phi|_{r=M}$ by integrating Eq.~(\ref{2nd}).
In a similar way, we can determine $\partial_r^n \phi|_{r=M}$.

\section{Regularity of Moncrief's variables}
\label{app:moncrief}

In this section we will explain Moncrief's parameterisation of metric and gauge field perturbations of the RN solution. The essential point is to expand each type of perturbation in a Regge-Wheeler basis for scalar, vector and tensor harmonics on $S^2$ which we have written covariantly in (\ref{S2scalar}, \ref{S2vector}, \ref{S2tensor}).

As explained Moncrief works in the Hamiltonian formalism. Hence the perturbation variables consist of the various metric and gauge field components and their conjugate momenta. We will need explicit expressions for $\delta \pi^{ij}, \delta E^i$. The former can be computed using~\cite{Moncrief:1975sb}
\be
\delta \pi^{ij} = \frac{\sqrt{g}}{2N} (g^{ik}g^{jl}-g^{ij}g^{kl})[ \partial_t \delta g_{kl} - \nabla_k \delta N_l -\nabla_l \delta N_k] \; ,   \label{pij}
\ee
where $\nabla_i$ is the connection associated to $g_{ij}$. The latter can be computed from
\be
\delta E^i = -\frac{\sqrt{g}}{N} g^{ij} \delta F_{tj} +\left( - \frac{\delta N}{N}+ \frac{1}{2} g^{kl}\delta g_{kl}\right) E^i -  g^{ik} E^l \delta g_{kl}   \label{ei} 
\ee
which is valid if the background solution satisfies $N_i=0, F_{ij}=0$, as is the case for the RN solution we will consider. Indeed the RN background is given by equations (\ref{ADMRN1}) and (\ref{ADMRN2}).

We will also show that the subset of Moncrief's variables which appear in the reduced gauge invariant perturbation equations, see table \ref{table}, are smooth on the horizon.

\subsection{Odd perturbations}

For odd perturbations Moncrief's parameterisation of the perturbed metric and gauge field is~\cite{Moncrief:1974am, Moncrief:1974gw}
\bea
\delta g_{ij} dx^i dx^j &=&2 h_1 dr \hat{e}_a dx^a + h_2 \hat{e}_{ab}dx^a dx^b \\
\delta N_i dx^i &=& h_0 \hat{e}_a dx^a  \\
\delta A_i dx^i &=& A \hat{e}_a dx^a  \; .
\eea
The momenta conjugate to $\delta g_{ra}=h_1 \hat{e}_a$ and $\delta g_{ab}= h_2 \hat{e}_{ab}$ are given by (see e.g.~\cite{Moncrief:1974am})
\be
\delta \pi^{ra}= \frac{p_1\sin \theta \hat{e}^a}{2l(l+1)} \;, \qquad  \qquad \delta \pi^{ab} = \frac{2p_2 \sin \theta  \hat{e}^{ab}}{l(l+1)(l-1)(l+2)}
\ee
where the normalisations are chosen so that $p_1,p_2$ is conjugate to $h_1,h_2$ when one integrates the action over $S^2$.  The momentum conjugate to $\delta A_a = A \hat{e}_a$ is
\be
\delta E^a = \frac{E \sin \theta \hat{e}^a }{l(l+1)}  \; .
\ee
Hence perturbations are described by the set of function $(A, E, h_1, h_2, p_1, p_2, h_0)$.

For $l>1$, Moncrief performs a canonical transformation to a new set of canonical variables $(h_1, h_2, A, p_1, p_2, E) \mapsto  (k_1, k_2, f_1, \pi_1, \pi_2, \pi_f)$, where in particular $\pi_f = A$ and $\pi_1=p_1$. All the new variables are gauge invariant except $k_2$ which is cyclic; its conjugate momentum $\pi_2$ is the only constraint function (which follows by variation with the respect to the lapse $h_0$). Hamilton's equations for the remaining pairs $(k_1, \pi_1)$ and $(f_1, \pi_f)$ can be combined to give two second order coupled equations for $\pi_f$ and $\pi_g= r \pi_1 - 2Ql(l+1)\pi_f/r$. These equations can be decoupled into two wave equations of the form (\ref{genform}) for $\psi=P_{\pm}$ which are two certain constant linear combinations of $\pi_f, \pi_g$ (these are given in~\cite{bicak2} and we have written them in the extreme case in equation (\ref{Ppm1}) and (\ref{Ppm2})).

For $l=1$ there are no tensor perturbations and so we have $h_2=p_2=0$. Moncrief performs a canonical transformation $(h_1, A, p_1,  E) \mapsto  (k_1, f_1, \pi_1, \pi_f)$, where in particular $\pi_f= A$ and $\pi_1 = \tfrac{1}{2}r^2 p_1-2Q A$. Variation of the lapse $h_0$ in this case gives the constraint $\pi_{1,r}=0$, and together with the fact that $k_1$ is cyclic, this implies $\pi_1$ is a constant. Hamilton's equations for the remaining pair $(f_1, \pi_f)$ can be written as a single second order wave equation of the form (\ref{genformstatic}) for $\psi=P_f \equiv \pi_f- 2Q\delta a/r$ where $\delta a \equiv -\pi_1/(6M)$ is a constant by the previous remarks.

The above discussion shows that to examine smoothness of $P_{\pm}$ ($l>1$) and $\pi_f$ ($l=1$), we need to only examine smoothness of $A, p_1$ from the original variables. The latter can be computed using (\ref{pij}) and we find
\be
\label{p1}
p_1=l(l+1)\left( \partial_t h_1- \partial_r h_0 +\frac{2}{r} h_0 \right) \; ,
\ee
giving an explicit expression in terms of metric components.

A general odd perturbation of the metric regular on $\mathcal{H}^+$ in EF coordinates is
\be
\delta g_{\mu\nu} dx^\mu dx^\nu = 2 ( h_v dv + h_r dr) \hat{e}_a dx^a + h  \hat{e}_{ab} dx^a dx^b
\ee
where $h_r, h_v, h$ are smooth on the horizon $\mathcal{H}^+$. Converting to static coordinates gives
\be
\delta g_{\mu\nu} dx^\mu dx^\nu = 2 \left[ h_v dt+ \left(h_r + \frac{h_v}{F} \right) dr \right] \hat{e}_a dx^a + h  \hat{e}_{ab} dx^a dx^b \; .
\ee
Hence comparing to Moncrief gives
\be
h_0 = h_v \qquad h_1 = h_r + \frac{h_v}{F} \qquad h_2 = h
\ee
and so we deduce that $h_0, h_2$ are regular on the horizon, but  $h_1$ is not. The momentum conjugate to $h_1$ (\ref{p1}) in EF coordinates is therefore
\be
p_1=l(l+1) \left(\partial_vh_r- \partial_r h_v +\frac{2}{r} h_v  \right)
\ee
which is indeed smooth on the horizon.

A general odd perturbation of the Maxwell field regular on $\mathcal{H}^+$ in EF coordinates is
\bea
\delta F_{\mu\nu} dx^\mu dx^\nu = (f_{va} dv+f_{ra}dr) \wedge \hat{e}_a dx^a + fY \hat{\epsilon}_{ab} dx^a \wedge dx^b  \; ,
\eea 
where $f_{va}, f_{ra}, f$ are smooth. In static coordinates this becomes
\be
\delta F_{\mu\nu} dx^\mu dx^\nu = \left[ f_{va} dt+\left(f_{ra}+ \frac{f_{va}}{F} \right) dr \right] \wedge \hat{e}_a dx^a +fY \hat{\epsilon}_{ab} dx^a \wedge dx^b  \; .
\ee
Comparing to Moncrief gives  $f = - \frac{1}{2} l(l+1)A$, where we have used $d \hat{e} = \hat{\epsilon} \hat{\nabla}^2 Y = - l(l+1) Y \hat{\epsilon}$. Therefore we deduce that $A$ is indeed smooth on $\mathcal{H}^+$.

\subsection{Even perturbations}

Even perturbations are more complicated so we will not explain the derivation of the variables we need. The metric and gauge field are parameterised as~\cite{Moncrief:1974ng, Moncrief:1975sb}
\bea
\delta g_{ij} dx^i dx^j &=&F^{-1}H_2Y dr^2+ 2 h_1 dr \hat{f}_a dx^a + r^2(G \hat{f}_{ab}+K \hat{g}_{ab})dx^a dx^b \\
\delta N_i dx^i &=& H_1 Ydr+ h_0 \hat{f}_a dx^a  \\
\delta N&=& - \frac{1}{2} F^{1/2} H_0 Y  \\
\delta A_i dx^i &=& (a_1 + a_{2,r}) Y dr + a_2 \hat{f}_a dx^a \\
\delta A_t &=& a_t Y \; .
\eea
There are analogous expressions for the conjugate momenta $\delta \pi^{ij}$ and $\delta E^i$; the only one we will actually need is
\be
\delta E^r = f_1 \sin \theta Y  \; .
\ee
We can compute this explicitly using (\ref{ei})
\be
\label{f1}
f_1= - r ^2\int_{S^2} \delta F_{tr} Y_{lm}^* +Q \left(H_0-H_2 +2 K -l(l+1) G  \right)  \; .
\ee

For $l>1$ Moncrief transforms $(H_2,h_1, K, G)$ to the variables $(k_1, k_2, k_3, k_4)$ defined by
\bea
&&k_1 = K + r F G_{,r} -\frac{2}{r} F h_1  \qquad k_2= \frac{1}{2F}(H_2-K) - \frac{r}{2F} K_{,r}+ \frac{rF'}{4F^2}  K  \label{k1k2}\\
&&k_3 =G \qquad k_4 = h_1  \label{k3k4} \; .
\eea
Moncrief shows that $k_1,k_2$ are gauge invariant. He eventually managed to reduce the perturbation equations to a pair  of coupled second order wave equations of the form (\ref{genformstatic}) for the variables\footnote{In Moncrief's paper~\cite{Moncrief:1975sb} the variable $Q_1$ is called $Q$; we have renamed this to avoid confusion with the electric charge $Q$.}
\bea
&&Q_1\equiv \frac{q_1\sqrt{(l-1)(l+2)}}{\Lambda} \qquad \text{where} \qquad  q_1\equiv  4r F^2 k_2 + l(l+1) r k_1  \\
&&H \equiv \tilde{F}- \frac{2Q q_1}{\Lambda r}\qquad \text{where} \qquad  \tilde{F} \equiv f_1+ Ql(l+1) k_3
\eea
where $\Lambda = (l-1)(l+2)+6M/r- 4Q^2/r^2$. It turns out that $\tilde{F}$ is gauge invariant and therefore $(Q_1,H)$ are both gauge-invariant. The variables appearing in the decoupled equations in this case are certain constant linear combinations $R_\pm$ of $(Q_1,H)$. Therefore to examine smoothness of the final variables $R_{\pm}$ we need to examine smoothness only of $q_1, k_3, f_1$.

For $l=1$ we must have $G=0$. Moncrief reduces this system to a single wave equation of the form (\ref{genformstatic}) for the gauge-invariant variable $\psi= H$ where\footnote{Here $\hat{k}_2 = F k_2$ where $k_2$ is the $l=1$ variable Moncrief used~\cite{Moncrief:1975sb}: we avoided referring to this to avoid confusion with $k_2$ for the $l>1$ variable.}
\be
H \equiv f_1 - \frac{4Q}{\Lambda} \hat{k}_2 \qquad \text{where} \qquad  \hat{k}_2= FH_2 -F^{3/2}\partial_r( r F^{-1/2} K) + K - \frac{2Fh_1}{r}  \; .  \label{l1H}
\ee
Hence to examine smoothness of $H$ we need to examine only $f_1, \hat{k}_2$.

A general even perturbation of the metric regular on the future horizon is
\be
\delta g_{\mu\nu} dx^\mu dx^\nu = (h_{vv}dv^2+ 2h_{vr}dvdr+h_{rr}dr^2)Y+ 2 ( h_v dv + h_r dr) \hat{f}_adx^a + r^2(K \hat{g}_{ab}+ G  \hat{f}_{ab} )dx^a dx^b
\ee
where all component functions are smooth. Converting to static coordinates gives
\bea
\delta g_{\mu\nu} dx^\mu dx^\nu &=& \left[ h_{vv} dt^2 + 2dtdr \left( h_{vr}
+\frac{h_{vv}}{F} \right) + dr^2 \left( h_{rr}+\frac{2h_{vr}}{F}+ \frac{h_{vv}}{F^2} \right) \right]Y \nonumber \\ &&+  \left[ h_v dt+ \left(h_r + \frac{h_v}{F} \right) dr \right]   \hat{f}_adx^a + r^2(K \hat{g}_{ab}+ G  \hat{f}_{ab} )dx^a dx^b  \; .
\eea
Comparing to Moncrief's variables shows $K,G$ are the same and
\bea
&&h_0 = h_v \qquad h_1 = h_r + \frac{h_v}{F}  \\ &&H_0 = \frac{h_{vv}}{F}\qquad H_1= h_{vr}
+\frac{h_{vv}}{F} \qquad   H_2 =  F h_{rr}+2h_{vr}+ \frac{h_{vv}}{F}  \; .
\eea
For $l>1$ we deduce that the quantities $k_1, F^2 k_2, k_3$ given in (\ref{k1k2}), (\ref{k3k4}) are smooth on the horizon, and hence $q_1$ also is.  For $l=1$ we deduce that $\hat{k}_2$ given in (\ref{l1H}) is smooth on the horizon.

A general even perturbation of the Maxwell field in EF coordinates is
\bea
&&\delta F_{\mu\nu} dx^\mu dx^\nu =f_{vr} Y dv \wedge dr+  (f_{va} dv+f_{ra}dr) \wedge \hat{f}_a dx^a  
\eea
which in static coordinates becomes
\bea
&&\delta F_{\mu\nu} dx^\mu dx^\nu =f_{vr} Y dt \wedge dr+  \left[ f_{va} dt+\left(f_{ra}+\frac{f_{va}}{F} \right)dr \right] \wedge \hat{f}_a dx^a  \; .
\eea
Therefore we find that (\ref{f1}) is given by
\be
f_1 = -r^2 f_{vr}+ Q( -F h_{rr}- 2 h_{vr} + 2 K -l(l+1) G  ) 
\ee
and hence this quantity is also smooth on the horizon.

Putting everything together we deduce: the gauge-invariant functions $R_\pm$  for $l>1$ and  $H$ for $l=1$ are indeed smooth on $\mathcal{H}^+$.

\end{document}